\documentclass[12pt,preprint]{aastex}
\newcommand\simlt{\lower.5ex\hbox{$\; \buildrel < \over \sim \;$}}
\newcommand\simgt{\lower.5ex\hbox{$\; \buildrel > \over \sim \;$}}

\slugcomment{To appear in the {\it Astrophysical Journal}}
\begin{document}

\title{Theory and astrophysical consequences of a magnetized 
           torus around a rapidly rotating black hole}
\author{Maurice H.P.M. van Putten}
\affil{LIGO Project, Massachusetts Institute of Technology, NW17-161, 
       175 Albany St., Cambridge, MA 02139-4307}
\author{Amir Levinson}
\affil{School of Physics and Astronomy, Tel Aviv University, Tel Aviv, Israel} 
%        {{\bf DRAFT} NOT FOR DISTRIBUTION}

\begin{abstract}  
   We analyze the topology, lifetime, and emissions of a  
   torus around a black hole formed in hypernovae and black hole-neutron star 
   coalescence. The torus is ab initio uniformly magnetized, represented by two 
   counter oriented current-rings, and develops a state of suspended accretion 
   against a ``magnetic wall" around the black hole. Magnetic stability of the 
   torus gives rise to a new fundamental limit ${\cal E_B}/{{\cal E}_k}<0.1$ 
   for the ratio of poloidal magnetic field energy-to-kinetic energy, corresponding
   to a maximum magnetic field strength $B_c\simeq 10^{16}\mbox{G}\left({7M_\odot}/{M_H}\right)
   \left({6M_H}/{R}\right)^2\left({M_T}/{0.03M_H}\right)^{1/2}$.  
   The lifetime of rapid spin of the black hole is effectively defined by the timescale of 
   dissipation of black hole-spin energy $E_{rot}$ in the horizon, and satisfies
   $T\simeq 40\mbox{s} (M_H/7M_\odot)(R/6M_H)^4(0.03M_H/M_T)$ for a 
   black hole of mass $M_H$ surrounded by a torus of mass $M_T$ and radius $R$.
   %The suspended accretion state may hereby be intermittent on an intermediate time-scale
   %of seconds, associated with magnetic field build-up powered by the rotational energy
   $E_{rot}$ of the black hole. The torus converts a major fraction 
   $E_{gw}/E_{rot}\sim 10\%$ into gravitational radiation through a finite 
   number of multipole mass-moments, and a smaller fraction into MeV neutrinos 
   and baryon-rich winds. At a source distance of 100Mpc, these emissions over 
   $N=2\times 10^4$ periods give rise to a characteristic
   strain amplitude $\sqrt{N}h_{char}\simeq 6\times 10^{-21}$.
   We argue that torus winds create an open magnetic 
   flux-tube on the black hole, which carries a minor fraction 
   $E_j/E_{rot}\simeq 10^{-3}$ in baryon-poor outflows to infinity. We conjecture 
   that these are not high-sigma outflows owing, {in part}, to magnetic reconnection in 
   surrounding current sheets. The fraction $E_j/E_{rot}\sim (1/4)(M_H/R)^4$ is 
   standard for a universal horizon half-opening angle $\theta_H\simeq M_H/R$ of the 
   open flux-tube. We identify this baryon poor output of tens of seconds with GRBs
   with contemporaneous and strongly correlated emissions in gravitational radiation,
   conceivably at multiple frequencies. Ultimately, this leaves a
   black hole binary surrounded by a supernova remnant.
\end{abstract}

%\keywords{black hole physics --- gamma-rays: bursts and theory -- gravitational waves}

%\mbox{}\\
%\newpage
\section{Introduction}

  Black holes surrounded by a magnetized torus or disk are believed to constitute 
  the central engines that power various high-energy sources, 
  notably active galactic nuclei, galactic microquasars, and gamma-ray bursts.  
  The latter systems are thought to be the outcome of catastrophic
  events such as core-collapse in massive stars and black hole-neutron star
  coalescence \citep{eich89,woo93,pac91,pac98}, and are of interest as
  potentially the most extreme and short-lived black hole-torus systems. 
  
  We describe a theory for the topology, lifetime and
  emissions of a black hole-torus sytems as a
  function of three parameters: the mass $M_H$ of an extreme Kerr black hole,
  the radius $R$ and the mass $M_T$ of the torus. 
  We shall do so largely by {studying the torus} by equivalence to pulsars,
  {in both topology and ms rotation periods.}
  The energy emissions are powered by the spin-energy of a Kerr black hole
  \citep{ker63}. 
  Most of the black hole-luminosity {-- the rate at which the black hole deposits
  energy into its surroundings in all channels --} is incident on the torus,
  which hereby creates a {\em major} energy output of the system.
  A {\em minor} energy output is released 
  in baryon-poor outflows through an open magnetic flux-tube
  along the spin-axis of the black hole. The life-time of these black hole-torus
  systems is identified with the lifetime of rapid spin of the black hole in a
  state of suspended accretion \citep{mvp01A}
  against a ``magnetic wall" around the black hole \citep{mvp99}.
  The suspended accretion state results from a
  strong coupling of the torus to the spin-energy of the black hole. We point
  out that this mechanism is
  based on a uniform magnetization of the torus, represented by two 
  oppositely oriented current rings \citep{mvp99}. This 
  magnetization is a natural 
  outcome of both black hole-neutron star coalescence and core-collapse 
  in hypernovae. 

  In this paper, we quantify (1)  
  {the lifetime of rapid spin of the black hole in terms of a new magnetic stability 
  criterion for the torus};
  (2) baryon-rich outflows from the torus at MeV temperatures; and
  (3) the fraction of black hole spin-energy in baryon-poor outflows
  through an open magnetic flux-tube on the black hole, created from
  outer layers of the inner torus magnetosphere by these torus winds.

  The torus is luminous in various channels, which are strongly correlated by the
  properties of the torus: in gravitational radiation, winds, thermal and MeV neutrino emissions 
  \citep{mvp01a,mvp02}. Calorimetric constraints on the torus winds hereby
  obtains predictions for the proposed emissions in gravitational radiation, while 
  calorimetry on the gravitational wave emissions by upcoming gravitational
  wave-experiments obtains a method for identifying Kerr black holes 
  as objects in nature \citep{mvp02}.

  Gravitational radiation forms a major output of the system and 
  the dominant output of the torus \citep{mvp01a,mvp02}. 
  This is emitted by a finite number of multipole mass moments
  in a torus of finite slenderness, due to the Papaloizou-Pringle instability 
  \citep{pap84,mvp02c} and, conceivably, other wave-modes in the torus. 
  These emissions are candidate sources for the upcoming 
  gravitational wave-experiments by laser {interferometric instruments 
  LIGO \citep{abr92}, VIRGO \citep{bra92}, TAMA \citep{mas01} and GEO 
  (e.g., \cite{geo}),} 
  or by any of the 
  bar or sphere detectors presently under construction. The frequency in 
  gravitational radiation is determined by the keplerian frequency of the
  torus. The latter is strongly correlated to the output energy in torus winds.
  This suggests the possibility of performing calorimetry on the impact of these
  torus winds on the remnant stellar envelope \citep{mvp02d} and on
  hypernova remnants, in order to constrain the expected frequency in 
  gravitational radiation.

  Baryon-poor outflows form a small fraction of the total output from the black hole
  through an open magnetic flux-tube \citep{mvp02}. We here attribute
  the formation of this open flux-tube to powerful baryon-rich torus winds, 
  driven from the surface by escaping MeV neutrinos. This neutrino output provides
  the dominant cooling mechanism of the torus during the suspended accretion state,
  in addition to the energy release in gravitational waves. 
  The fraction of rotational energy thus released is proportional to $\theta_H^4$,
  which is standard for a universal horizon half-opening angle $\theta_H$ 
  of the inner tube. These outflows are probably not high-sigma, owing to magnetic 
  reconnection in the interface between the baryon poor and baryon-rich winds in a 
  surrounding outer flux-tube. The half-opening angle is possibly related
  to curvature in poloidal topology \citep{mvp02d}.

  Tentative observational evidence for a black hole-luminosity incident into
  surrounding matter in case of supermassive black holes is found in
  MCG-6-30-15 \citep{wil01}, and in case of stellar mass black hole-candidates in
  the galactic source XTE J1650-500 \citep{mil02}. Independent
  observational evidence of the presence of magnetic fields remains elusive. We
  point out, however, that in our model heating of the torus is due to
  viscous shear between its inner and outer faces, 
  possibly in the form of magnetohydrodynamical turbulence, which is different
  from electric dissipation as envisioned in \cite{wil01}.

  In \S\ref{sec:T-magnetosphere} and \S\ref{sec:stability} we describe the topology 
  and stability of the magnetosphere of a torus formed in black hole-neutron star
  coalescence and hypernova, the suspended accretion state around rapidly rotating 
  black holes and the secular time-scale {of its rapid spin}. The formation of
  a finite number of multipole mass moments by the Papaloizou-Pringle instability
  in tori of finite slenderness is summarized in \S\ref{sec:susaccrt}. 
  Energy emissions in various channels by the torus are described in \S\ref{sec:dissp}. 
  We calculate the neutrino driven mass loss rate from 
  the torus' surface in \S\ref{sec:massejection}. These baryon-rich torus winds 
  have some implications for 
  energy extraction from the black hole and the creation of open magnetic flux-tubes
  from outer layers in the inner torus magnetosphere. \S \ref{sec:opentubes}- 
  \S \ref{Sec:interface} describe the structure
  of the inner flux-tube supported by the black hole, and dissipation by 
  magnetic reconnection in its interface with the surrounding outer flux-tube.
  We conclude with observational consequences in \S\ref{Sec:conclusion}.

\section{Formation and structure of the torus magnetosphere}
% of a force-free torus magnetosphere}
\label{sec:T-magnetosphere}
\begin{figure}
%\plotone{f2002d}
%\plottwo{c_A}{c_B}
\plotone{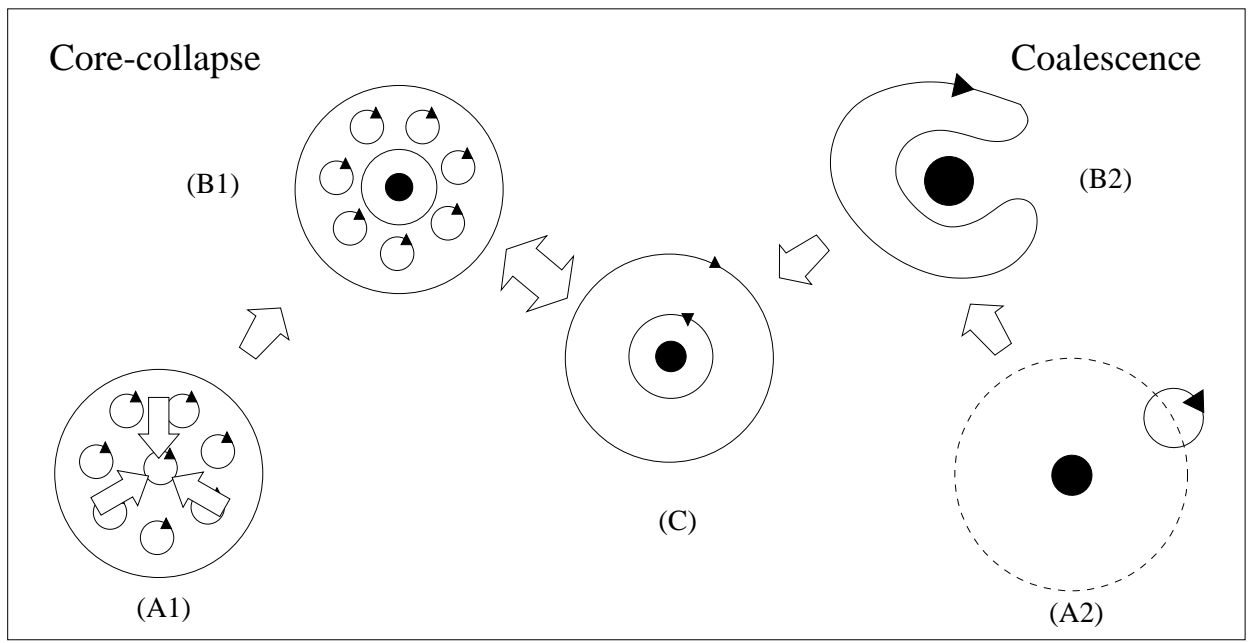}
\caption{A uniformly magnetized torus around a black hole (C) is  
represented by two counter-oriented current rings in the equatorial plane. It forms
a common end point of both core-collapse (A1,B1,C) and black hole-neutron star
coalescence (A2,B2,C). Core-collapse (A1-B1) in a magnetized 
star results in a uniformly magnetized, equatorial annulus (C); tidal break-up 
(A2-B2) wraps the current ring representing the magnetic moment of a neutron star 
around the black hole which, following a reconnection, leaves the same (C).}
\end{figure}
A torus formed in core-collapse of a massive star or black hole-neutron
star coalescence will be magnetized with a remnant of the progenitor magnetic field. 
An aligned poloidal magnetic field in the progenitor star provides a magnetic moment 
density in the torus, aligned with its axis of rotation. Equivalently, the
magnetic field in the torus is produced by two concentric current loops with
opposite orientation. In the case of a torus formed from the break-up of a
neutron star around a black hole, these two current loops form out of a single
current loop representing the magnetization of the neutron star, upon stretching
the latter around the black hole followed by a reconnection (see Fig. 1). 
Conceivably, the magnetic field in the torus is amplified by winding or a 
dynamo process. The stability of this magnetic field configuration is discussed
in \S \ref{sec:stability}, and sets an upper limit of 
$B\simeq10^{16}$G on the field-strength.

Below, we outline the resulting magnetosphere of the black hole torus 
system in the force-free limit.  For illustrative purposes we consider first 
the poloidal topology of the vacuum magnetic field configuration.  We proceed by
discussing the more realistic situation of a force-free magnetosphere, and show
that the the inner and outer face of the torus are each equivalent to a pulsar
with, however, generally different angular velocities. 

In the subsequent sections we shall consider some essential energetic aspects in 
greater detail.  In particular, it will be shown that 
{appreciable} matter outflows
are expected along open field lines to infinity, and a reconnection 
boundary is expected to form near the rotation axis.  This may give rise 
to a non-force-free magnetic field in those regions.

\subsection{Topology of the vacuum magnetic field}
\begin{figure}
%\plotone{pp200}
\plotone{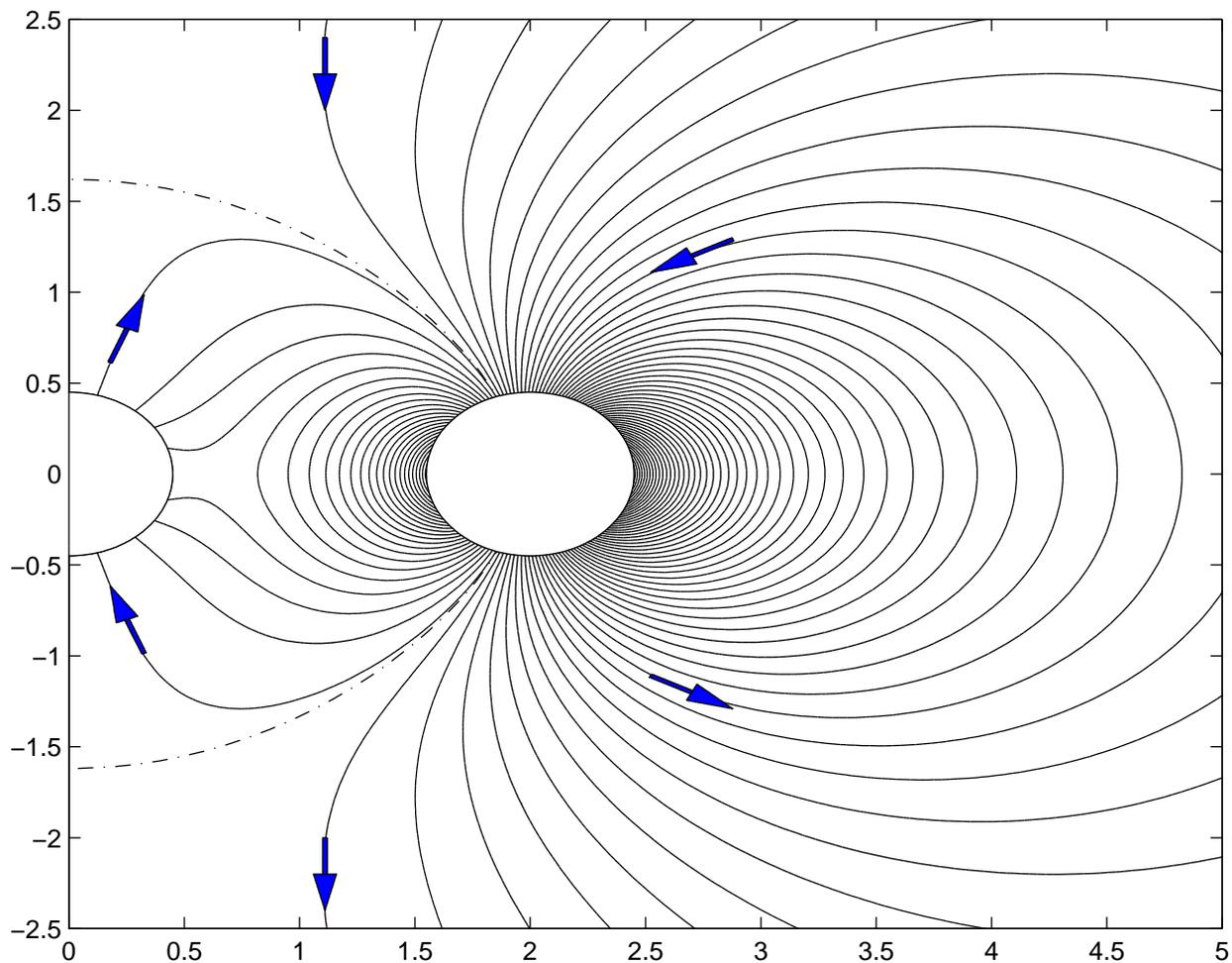}
\caption{The poloidal topology of magnetic flux-surfaces in vacuum is  
{illustrated in flat space-time,}
produced by two counter-oriented current rings representing a uniformly magnetized
torus ({\em center}). The dashed line is the separatrix between the flux-surfaces
supported by the inner and the outer faces of the torus. 
{A Kerr black hole develops an equilibrium magnetic moment which preserves
essentially uniform and maximal magnetic flux through its horizon.}}
\end{figure}
A uniform magnetization of the torus is approximately described by 
two counter-oriented current rings in the equatorial plane.  A third current
loop is associated with the black hole, representing its equilibrium magnetic
moment in {its lowest energy state \citep{mvp01a}.} 
{This induced magnetic moment is oriented antiparallel to the magnetic
moment of the torus}, facilitating an essentially uniform and
maximal horizon flux at arbitrary spin-rates.
In flat space-time, the magnetic field produced by a superposition of current
rings can be calculated analytically \citep{jac75}.
The topology of this ab initio flat-spacetime vacuo magnetic field
is shown in Fig. 2.  As seen, at large radii (compared with the 
radius of the outer current ring) it quickly approaches a dipole solution.  
In the inner region the field lines intersect the 
horizon, giving rise to a strong coupling between the black hole and the inner
face of the torus. This topology and flux-distribution is preserved in the face
of general relativistic effects, as a result of the equilibrium magnetic moment of
the black hole.

\subsection{Equivalence to pulsars}
By vacuum break-down, the flux-surfaces will evolve with electric charges to a
largely force-free state similar to pulsars \citep{gol69}. 
As a result, a magnetosphere develops which
consists of conductive flux-surfaces and magnetic winds.
The torus hereby supports an inner 
and an outer torus magnetosphere. These are equivalent in poloidal topology
to pulsar magnetospheres, 
wherein the horizon of the black hole is equivalent to a compactified infinity with 
non-zero angular velocity (Fig. 3). This equivalence implies similar, causal 
interactions by magnetic winds acting on the inner and outer face of the torus.

\begin{figure}
%\plotone{f2002d}
%\plottwo{c_A}{c_B}
\plotone{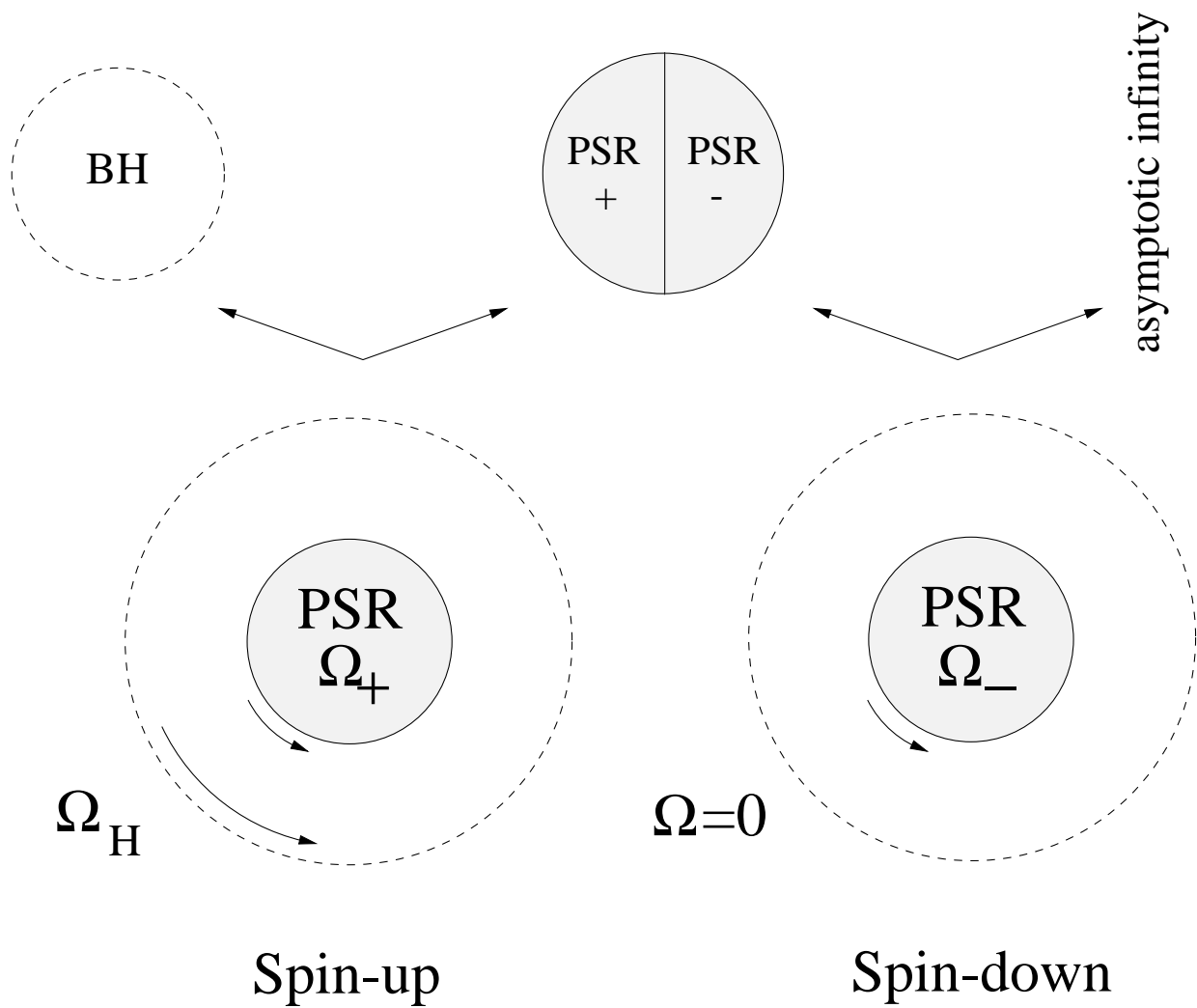}
\caption{{\em Lower left.} The inner face of the torus (angular velocity $\Omega_+$) 
and the black hole (angular velocity $\Omega_H$) is equivalent to a pulsar
surrounded by infinity with relative angular velocity $\Omega_H-\Omega_+$.
By equivalence in poloidal topology to pulsar magnetospheres, the inner
face receives energy and angular momentum from the black hole as a causal process, 
whenever $\Omega_H-\Omega_+>0$. {\em Lower right.}
The outer face of the torus (angular velocity $\Omega_-$)
is equivalent to a pulsar with angular velocity $\Omega_-$, and always looses
energy and angular momentum, by the same equivalence.}
\end{figure}

In the force-free limit, the flux-surfaces in the outer/inner torus magnetosphere  
assume rigid corotation with the outer/inner face of the torus, by
no-slip boundary conditions on the surface. Field-lines in the vicinity of the
torus and the equatorial plane form a `bag' of closed field-lines (both on the inner
and the outer face) with no-slip/no-slip boundary conditions. The last closed 
field-line of the outer torus magnetosphere reaches the light-cylinder associated 
with the angular velocity of the outer face of the torus, similar to the last closed 
field-lines in pulsar magnetospheres. The last closed field-line of the 
inner torus magnetosphere reaches the inner light-surface \citep{zna77} associated with the 
angular velocity of the inner face of the torus (Fig. 2 of \cite{mvp99}). 
Beyond, field-lines are open, and
extend to infinity or to the horizon of the black hole with no-slip/slip boundary
conditions. 
The former are created by torus winds which cross the outer light cylinder.

%Torus winds may further create open field-lines to infinity from the inner torus
%magnetosphere by folding of magnetic flux-surfaces, leaving an open flux-tube 
%supported by the black hole which extends to infinity. 

\subsection{Suspended accretion against a magnetic wall}

Most of the black hole-luminosity is incident on the torus \citep{mvp99}.
The torque exerted on the inner face of the torus by the black hole is 
obtained by integrating the angular momentum flux ${\cal L}^{r}=
F^{r\theta}F_{\phi\theta}/4\pi$ 
over the section of the horizon which is threaded by magnetic field
lines that are anchored to the torus:
\begin{equation}
T_{+}=4\pi\int_{\theta_H}^{\pi/2}{\sqrt{-g}{\cal L}^{r}d\theta}=
(\Omega_H-\Omega_{+})\int_{\theta_H}^{\pi/2}\frac{\Sigma}{\rho^2}
\sin\theta(F_{\phi\theta})^2 d\theta,
\label{torque}
\end{equation}
where eq. (\ref{ratio1}) has bee used.  Here $\theta_H$ is the angle of the 
last field line that connect the torus and the horizon (see Fig. 3), $\Omega_{+}$ and
$\Omega_{H}$ are the angular velocities of the inner face of the torus 
and the black hole,
respectively, and the metric components, $\rho$, $\Sigma$ are defined in 
the appendix.  In terms of the net poloidal magnetic flux associated with
the open field lines in the torus, $2\pi A$, we may write 
$T_{+}=(\Omega_H-\Omega_{+})f_H^2A^2$.  Formally $f_{H}$ is defined through 
eq. (\ref{torque}).  Likewise, the torque exerted on the outer face by field lines
that extend to infinity can be expressed as: $T_{-}=\Omega_{-}f_{w}^2A^2$,
where $\Omega_{-}$ is the angular velocity of the outer face of the torus,
and $f_{w}$ is the fraction of open field lines that extend to infinity.
It is seen that when $\Omega_H>\Omega_{+}$ angular momentum is transferred 
from the black hole to the torus, tending to spin up the inner face, whereas 
in the slowly rotating case ($\Omega_H<\Omega_{+}$) the black hole
receives angular momentum from the torus (Fig. 4).
\begin{figure}
\plotone{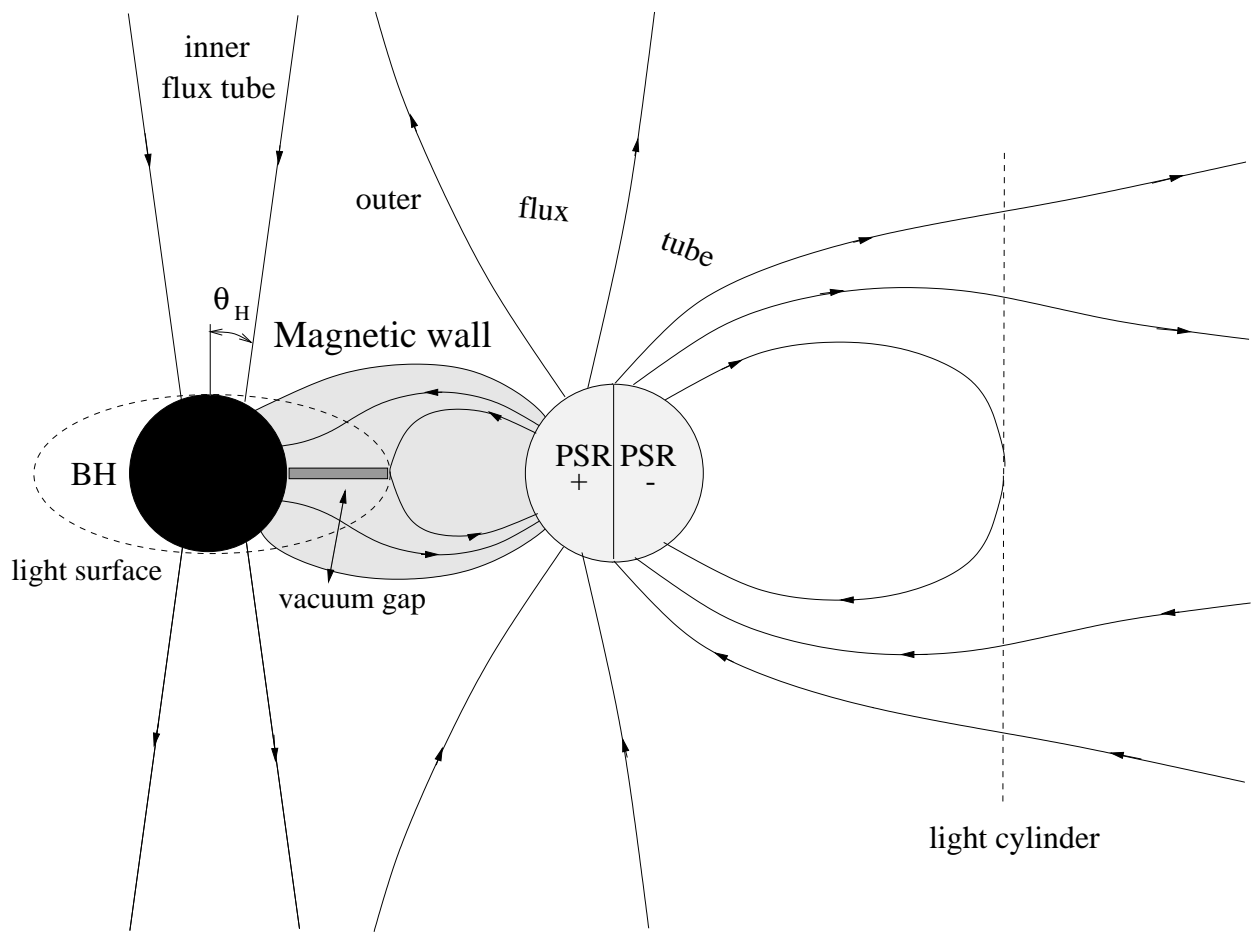}
%%%\centerline{\plotone{T2.xfig.eps}}
%\centerline{\plotone{f2002c.eps}}
%\centerline{\plotone{f2002a.eps}}
%\centerline{\plotone{./BE-topol.eps}}
%\epsfig{file=BE-topol.eps,width=170mm,height=200mm}
\vskip0.1in
\caption{Schematic illustration of the poloidal topology of the magnetosphere 
of a torus surrounding a rapidly rotating black hole. Balance of the input on
the inner face and the output through the outer face and other losses results
in a suspended accretion state. The inner torus magnetosphere hereby represents
a magnetic wall around the black hole, whereby the torus receives energy and
angular momentum from the black hole by equivalence to pulsars. The associated
horizon Maxwell stresses (Blandford \& Znajek 1977) and Maxwell stresses on the
inner face of the torus (van Putten 1999) are mediated by poloidal currents, which
close over a vacuum gap in an annulus of vanishing magnetic field.
The torus hereby receives a {\em major} fraction of the black hole-spin energy,
catalyzing this into gravitational radiation, winds, thermal 
and MeV neutrino emissions. A baryon-poor inner flux-tube serves as an artery for 
a {\em minor} fraction of black hole-spin energy. The dashed line
indicates the light cylinder of the outer face of the torus.
The lifetime of the system is set by the lifetime of rapid spin of the black hole.}
\end{figure}
The outer face always loses angular momentum via a wind to infinity.  
{By mechanical work},
the magnetic torus winds to infinity and into the horizon carry
outgoing luminosities 
\begin{equation}
L_{\pm}=\Omega_{\pm}T_{\pm}.
\label{Lpm}
\end{equation}
The inner face thus receives a fraction $\Omega_{+}/\Omega_H$ of the maximum
extractable power $\Omega_HT_{+}$ (Thorne et al. 1986); the rest dissipates 
on the horizon.  

The positive and negative torques exerted on the inner and outer faces of the
torus in the rapidly rotating case, give rise to a differential rotation of 
the torus.  In steady state, a flow of angular momentum from the inner to the outer 
face is accomplished through shear forces.  The nature of the latter is
presumably electromagnetic \citep{bh92}, stimulated by magneto-hydrodynamic 
instabilities (see \S4).  As a consequence, a fraction 
of the black hole spin down energy that is intercepted by the inner face 
is dissipated in the torus, heating it to MeV temperatures. 

\subsection{Frame-dragging and electric charge distribution}
\begin{figure}
%\label{Fig:charge}
\centerline{\plotone{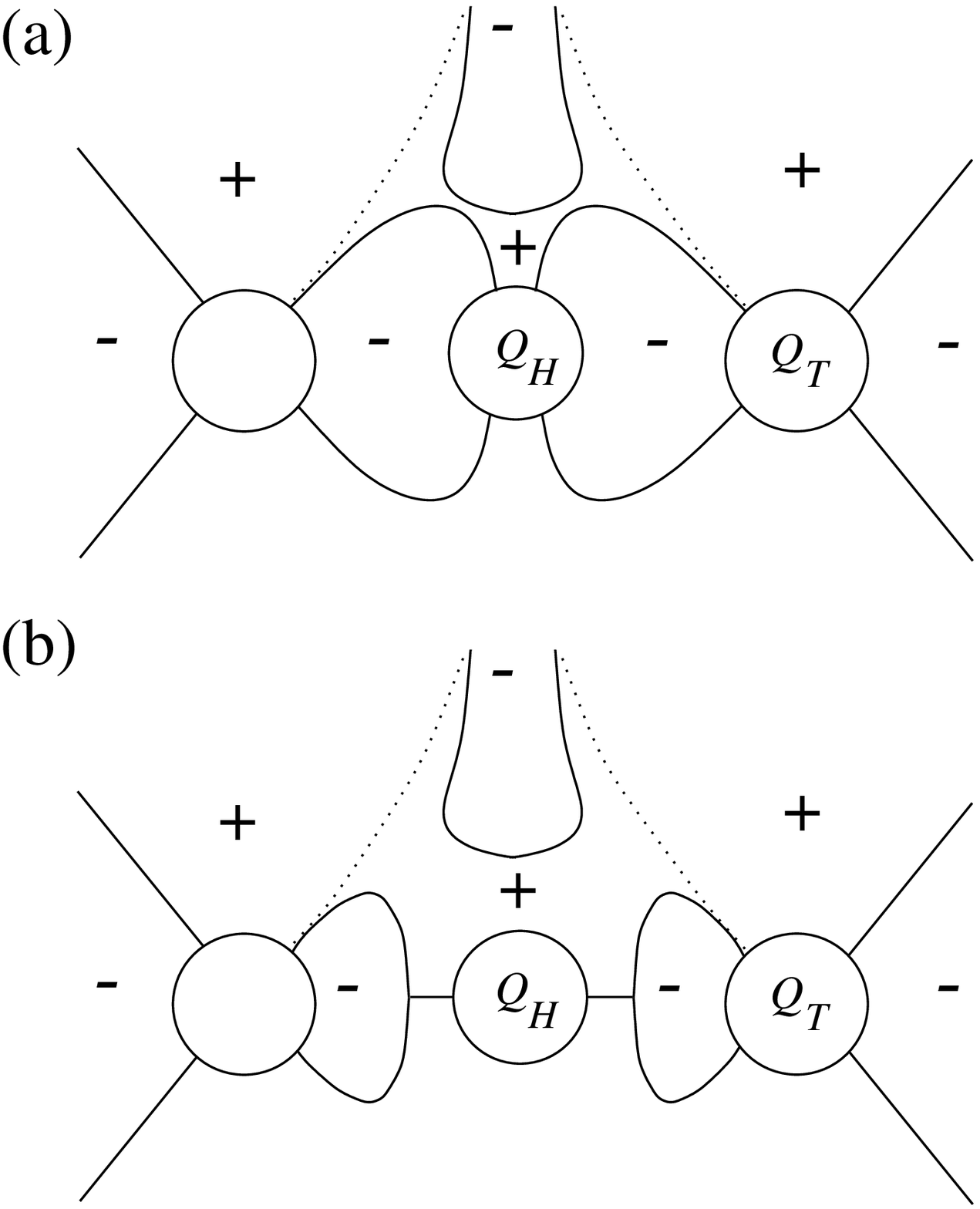}}
\vskip0.1in
\caption{Shematic of the charge-distribution in a torus 
magnetosphere around Kerr black holes, showing two bifurcations from 
non-rotation (dotted line). 
Frame-dragging introduces a sign-change in flux-tubes which rotate
slower than the black hole (a).
Black holes rotating faster than the torus produce
an inner light surface, and hence an annulus of $B=0$ (b).
Equilibrium magnetic moments of the black hole and the torus correspond to charges
$Q_H$ and $Q_T$ with opposite sign ($Q_T$ is analogous to 
pulsar charges \cite{coh75}).}
\end{figure}
The electric charge distribution
can be obtained from Maxwell's equation: $F^{t\mu}_{\ \ ;\mu}=4\pi j^{t}$.  
Assuming the radial magnetic field, $B_r=-F_{\phi\theta}/(\Sigma\sin\theta)$, 
to be independent of $\theta$, we obtain from eq. (\ref{Max-t}) the 
Goldreich-Julian charge density near the horizon, on field lines that 
emanate from the inner face:
\begin{equation}
\rho_e=\alpha^2 j^t=-\frac{(\Omega_{+}+\beta)B_r\cos\theta}{2\pi},
\label{EQN_RHO}
\end{equation}
where $\beta$ is defined below eq. (\ref{metric}), and equals $-\Omega_H$ 
on the horizon.  We find that for $\Omega_H>\Omega_{+}$, the charge density 
changes sign on field lines threading the horizon.  The same holds true 
also for the inner flux tube that extend to infinity 
(see discussion following eq. [\ref{rho_e}]).

The charge-distribution (\ref{EQN_RHO}) 
represents a bifurcation from the magnetosphere around
a Schwarzschild black hole. Ingoing horizon boundary conditions and outgoing
boundary conditions at infinity may hereby carry a continuous electric current.
If the black hole rotates slower than the inner face of the torus, the inner 
light-surface is absent and the sign of the charge-distribution near the torus
carries through to the horizon (Fig. 5a).  
If the black hole rotates faster than the inner face of the torus, the inner
light-surface becomes apparent (a bifurcation from slow rotation) and, 
by forgoing arguments, introduces a sign change in the charge-distribution
(Fig. 5b).

\section{The lifetime of rapid spin of the black hole}
\label{sec:stability}

A magnetized torus of a few tenths of solar masses around a stellar mass
black hole of about $7M_\odot$ is subject to magnetic self-interaction 
and a stabilizing tidal interaction in the central potential well. We here
derive limits on the average strength of a poloidal magnetic field which can
be supported by the torus. We do not consider the problem of stability of
a poloidal magnetic field itself, such as the 
{magnetorotational} (MRI) instability.
{A limit on the energy in poloidal magnetic field defines a lower
bound on the lifetime of rapid spin of the black hole.}

In regards to wave-motion within the equatorial plane, the contribution of
poloidal magnetic fields is that of magnetic pressure, which is generally
stabilizing on the motion of the fluid. In regards to poloidal wave-motion,
a poloidal magnetic field generally conspires towards instabilities. This
can be calculated by partitioning the torus in a finite number of fluid 
elements with current loops, representing local magnetic moments. 
The two leading-order partitions are shown in configurations C and B1 of 
Fig. 1, for which we derive critical magnetic field-strengths. The first
is subject to magnetic tilt instability between the inner and the outer 
face, and the second is subject to a magnetic buckling instability.
\begin{figure}
\plotone{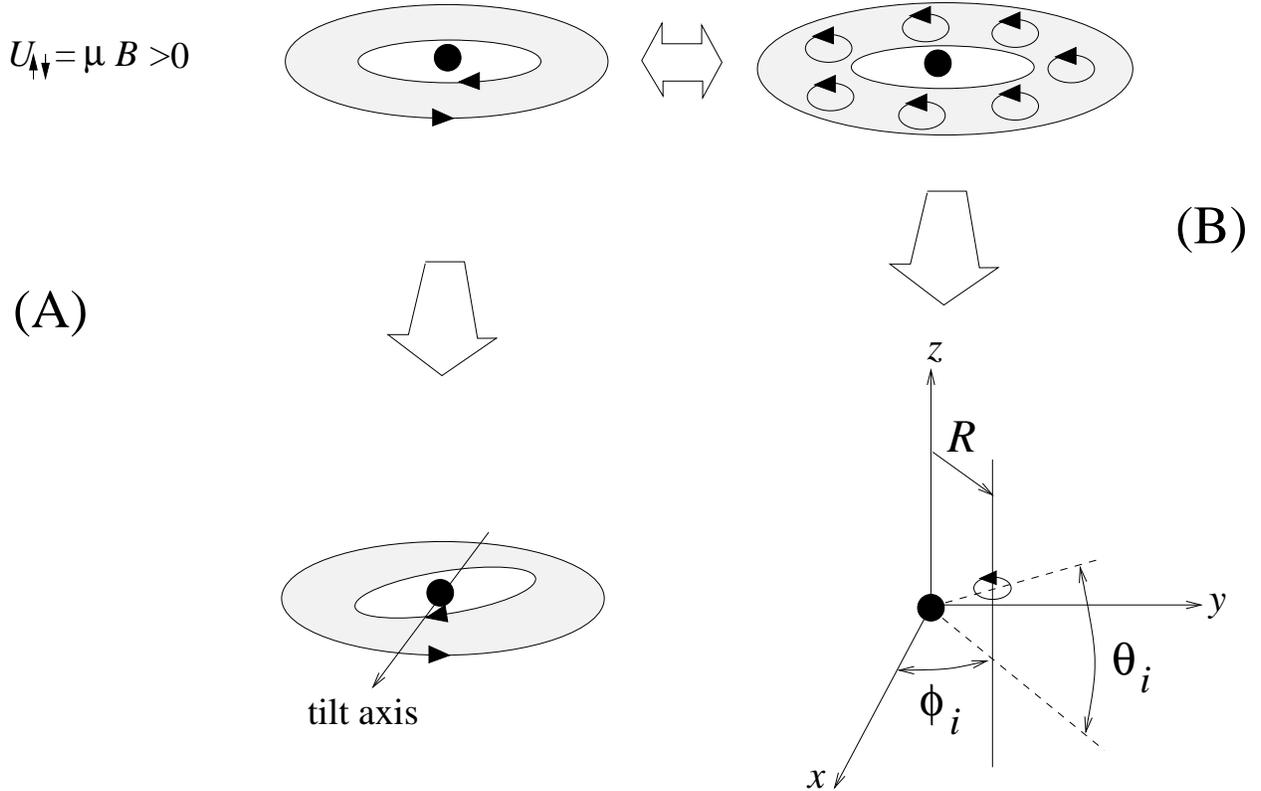}
\caption{A uniformly magnetized torus is in its highest magnetic energy
state. The two alternative leading-order partitions of the current 
distribution ({\em top}) have unstable poloidal modes, described by a
relative tilt between the two current rings towards alignment (A) or 
towards buckling (B), 
characterized by perturbations out of the equatorial plane with
poloidal angles $\theta_i\ne\theta_j$.
We consider vertical displacements of fluid elements along a cylinder of radius $R$.
Their wave-modes are stabilized by the tidal gravitational field
of the central black hole, provided the magnetic field-strength remains 
below a critical value on the order of $10^{16}$G. This gives 
rise to a minimim lifetime of rapid spin of the black hole on the order
of tens of seconds, consistent with the redshift-corrected durations of 
long GRBs.}
\end{figure}

%\centerline{\em(a) A magnetic flip instability}
\subsection{A magnetic tilt instability}

Following C in Fig. 1, consider 
the magnetic interaction energy of a pair of concentric current rings, given by
\begin{eqnarray}
U_\mu(\theta)=-\mu B \cos\theta.
\end{eqnarray}
Here, $\mu$ is the magnetic moment of the inner ring, $B$ is the
magnetic field produced by the outer ring, and $\theta$ denotes the angle between 
$\mu$ and $B$. Note that $U_\mu(\theta)$ has a period $2\pi$, is maximal (minimal)
when $\mu$ and $B$ are antiparallel (parallel; see Fig. 6). Consider
tilting a fluid element of a ring out of the equatorial plane to a height $z$
approximately along a cylinder of radius $R$. 
(This is different from tilting a rigid ring,
whose elements move on a sphere). 
A tilt hereby changes the distance to central black hole to
$\rho=\sqrt{R^2+z^2}\simeq R(1+z^2/2R^2)$.
In the approximation of equal mass
in the inner and outer face of the torus, simultaneous tilt of one 
ring upwards and the other ring downwards is associated with the potential energy
\begin{eqnarray}
U_g(\theta)\simeq -\frac{M_{T}M_H}{R} \left(1-\frac{1}{4}\tan^2(\theta/2)\right),
\end{eqnarray}
with $\tan(\theta/2)=z/R$, where we averaged over all segments of a ring. 
Note that $U_g(\theta)$ has period $\pi$ and is minimal when $\theta=0$. 
Stability is accomplished provided that the total potential energy
$U(\theta)=U_\mu(\theta) + U_g(\theta)$ satisfies
\begin{eqnarray}
\frac{d^2U}{d\theta^2}>0.
\label{EQN_STAB}
\end{eqnarray}
The potential $U(\theta)$ is shown in Fig. 7, which shows the bifurcation
at $b=1/12$ of the stable equilibrium $\theta=0$ into an unstable equilibrium
with the appearance of two neighboring stable equilibria at non-zero angles.
The bifurcation point is therefore second order. Nevertheless, the torus may
become nonlinearly unstable at large angles $(b>>b^*)$. We therefore consider
below the physical parameters at this bifurcation point.

\begin{figure}
\plotone{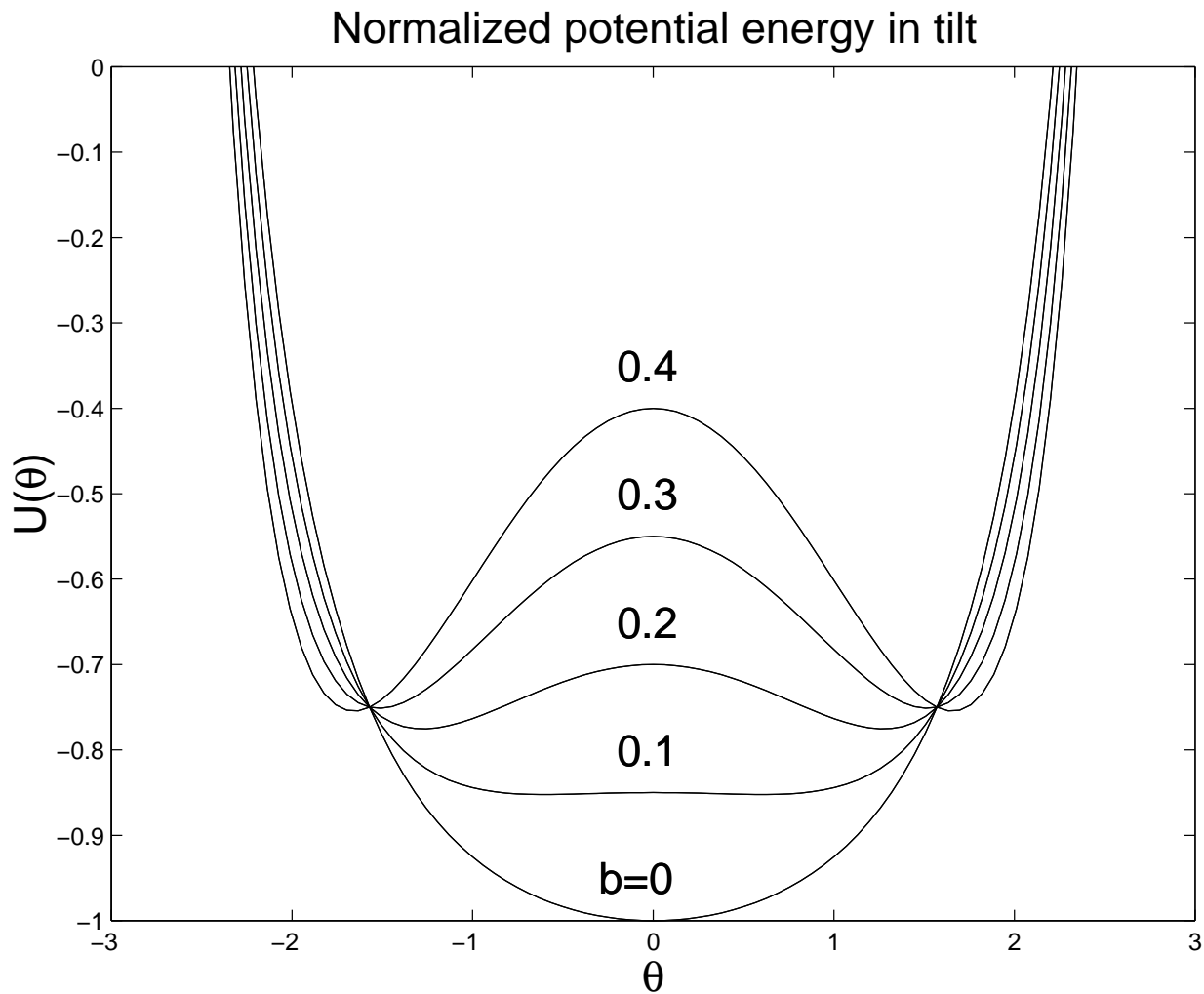}
\caption{The potential energy associated with a poloidal tilt angle
$\theta$ between the inner and the outer rings is the sum of a
magnetic moment-magnetic moment interaction plus a tidal interaction
with the central potential well of the black hole. It is shown for
various normalized magnetic field-energies
$b={\cal E}_B/{\cal E}_k$. 
The equilibrium $\theta=0$ becomes unstable when $d^2U(\theta)/d\theta^2<0$,
corresponding to a bifurcation into two stable branches of non-zero angles
beyond $b>1/12$.}
\end{figure}

For two rings of radii $R_\pm$ with $(R_+-R_-)/(R_++R_-)=O(1)$, we  
have $U_\mu\simeq \frac{1}{2}B^2R^3 \cos\theta$, so that (\ref{EQN_STAB})
gives
$B^2_cM_H^2=(1/4)({M_H}/{R})^4({M_T}/{M_H})$, or
\begin{eqnarray}
B_c\simeq 10^{16}\mbox{G}\left(\frac{7M_\odot}{M_H}\right)
\left(\frac{6M_H}{R}\right)^2\left(\frac{M_T}{0.03M_H}\right)^{1/2}.
\label{EQN_BC}
\end{eqnarray}
The critical value of the ratio of poloidal magnetic energy 
$({\cal E}_B=f_BB^2R^3/6)$ to kinetic energy $(M_TM_H/2R)$ in the torus
becomes
\begin{eqnarray}
\left.\frac{\cal E_B}{{\cal E}_k}\right|_c=\frac{f_B}{12},
\label{EQN_RC}
\end{eqnarray}
where $f_B$ denotes a factor of order unity, representing 
the volume of the inner torus magnetosphere as a fraction of
$4\pi R^3/3$. We emphasize that 
the limit (\ref{EQN_RC}) is fundamental, independent of the mass of the
black hole, and the mass and radius of the torus. 
This result can equivalently be attributed to stable 
         balance of Lorentz forces against tidal forces, preventing 
         a small misalignment to cause a tilt.

The poloidal magnetic
field introduces an anisotropic pressure tensor. At the critical magnetic
field-strength (\ref{EQN_BC}), pressure components in
the equatorial plane are predominantly magnetic rather than thermal
at MeV temperatures. The poloidal pressure components (along the magnetic
field-lines) are thermal pressures. In equilibrium, these poloidal pressure 
components are unaffected by poloidal tidal forces if the poloidal
flux-surfaces assume a spherical shape inside of the torus.

The magnetic field may be generated in response to the power received
from the black hole. If so, then
${d{\cal E_B}}/{dt}\le L_H$, where \citep{tho86,mvp99}
\begin{eqnarray}
L_H\simeq \frac{1}{32}\eta B^2M_H^2,
\label{EQN_LH}
\end{eqnarray}
denotes the black hole-luminosity expressed in terms of a fraction
$\eta=\Omega_T/\Omega_H$ of the angular velocity of the torus to
that of the black hole. The discrepancy $L_H-d{\cal E}_B/dt$ is 
carried away in the various channels as described in the suspended accretion
state, and will be zero in equilibrium. 
This gives rise to a minimum e-folding time 
\begin{eqnarray}
\tau_B=\frac{8}{3}\eta^{-1}\left(\frac{R}{M_H}\right)^2 R
      = 0.2\mbox{s} \left(\frac{\eta}{0.1}\right)^{-1}
                    \left(\frac{R}{6M_H}\right)^3
                    \left(\frac{M_H}{7M_\odot}\right).
\label{EQN_SUSPENDED}
\end{eqnarray}
A critical magnetic field of $B_c\simeq 10^{16}$G is hereby reached on a 
time-scale of at least a few seconds.
The suspended accretion state may hereby be intermittent on an intermediate time-scale
of seconds, associated with magnetic field build-up powered by the rotational 
energy of the black hole.

Most of the rotational energy is dissipated in the horizon, creating
entropy $S$ for a black hole temperature $T_H$ at a maximal rate
$T_H\dot{S}\simeq B^2M^2/32$ \citep{tho86}.
The lifetime of rapid rotation of the black hole becomes effectively the 
timescale of dissipation of black hole-spin energy $E_{rot}\simeq M_H/3$ in the horizon, 
i.e.:
\begin{eqnarray}
T\simeq\frac{E_{rot}}{T_H\dot{S}}\ge
40\mbox{~s}~\left(\frac{M_H}{7M_\odot}\right)
\left(\frac{R}{6M_H}\right)^4
\left(\frac{0.03M_H}{M_T}\right),
\label{EQN_LIFETIME}
\end{eqnarray}
where eq. (\ref{EQN_BC}) has been employed.  

{The suspended accretion state, conceivably
intermittent on a timescale (\ref{EQN_SUSPENDED}), lasts for
a duration similar to (\ref{EQN_LIFETIME}) until the ISCO reaches
the torus or until $\Omega_H\simeq\Omega_T$, whichever comes first.}

%\mbox{}\\
%\centerline{\em (b) A magnetic buckling instability}
\subsection{A magnetic buckling instability}

We partition the magnetization of the torus into $N$ equidistant
fluid elements with dipole moments, $\mu_i=\mu/N=(1/2)BR^3/N$.
We consider the vertical degree of freedom, of fluid elements which move 
to a height $z$ above the equatorial plane. By conservation
of angular momentum, this motion is restricted to a cylinder of
constant radius. 
Their position vectors in and off the equatorial plane will be denoted by
\begin{eqnarray}
{\bf r}_i^e=(R\cos\phi_i,R\sin\phi_i,0),~~~
{\bf r}_i  =(R\cos\phi_i,R\sin\phi_i,z_i),
 ~~~\phi_i=2\pi i/N.
\end{eqnarray}
A fluid element $i$ assumes an energy which consists of 
magnetic moment-magnetic moment interactions and the tidal
interaction with the central potential well. The total potential energy of
the $i-$th fluid element is given by
\begin{eqnarray}
U_i=-\frac{\mu_i B^\prime}{N}\Sigma_{j\ne i}
   \frac{|{\bf r}^e_i-{\bf r}^e_j|^3}{|{\bf r}_i-{\bf r}_j|^3}  
   \cos\theta_{ij} 
    + U_{g}(\theta_i),
\label{EQN_POT}
\end{eqnarray} 
where $B^\prime=B/N^*$ denotes the magnetic field-strength of a magnetic
dipole at distance $d=2\pi R/N$, $\theta_{ij}$ denotes the angle between the 
$i-$th magnetic moment
and the local magnetic field of the $j-$th magnetic moment, and
$U_{gi}=-(M_TM_H/RN)(1-z_i^2/2R^2)$ 
the tidal interaction of the $i-$th fluid element with the black hole.
Here, $N^*$ is a factor of order $N$ which satisfies the normalization
condition $\Sigma_i U_i=-\mu B$ (in equilibrium).
Upon neglecting azimuthal curvature in 
the interaction of neighboring magnetic moments, we have 
a magnetic moment-magnetic moment interaction
\begin{eqnarray}
{\mu_i B^\prime}\frac{|{\bf r}^e_i-{\bf r}^e_j|^3}{|{\bf r}_i-{\bf r}_j|^3} 
\cos\theta_{ij}\simeq
\frac{\mu_i B^\prime}
{|i-j|^3}\left(1-\left[1+\frac{3}{2|i-j|^2}\right]\alpha^2_{ij}\right),
\end{eqnarray}
where $\alpha_{ij}=(z_i-z_j)/d,$ 
%$d=2\pi R/N$, 
\begin{eqnarray}
\cos\theta_{ij}=-\sqrt{{(1-\alpha_{ij}^2)}/({1+\alpha_{ij}^2})}
\simeq -\left(1-\alpha_{ij}^2\right),
\end{eqnarray}
and $|{\bf r}_i-{\bf r}_j|\simeq |i-j|d
\left(1+{\alpha_{ij}^2}/{2|i-j|^2}\right).$ 
%\end{eqnarray}

We shall use the small a small amplitude approximation, whereby
$z_i/R=\tan\theta_i\simeq \theta_i$. We study the stability of this
configuration, to derive an upper limit for the magnetic field-strength.
An upper limit obtains by taking into account only interactions between
neighboring magnetic moments. (The sharpest limit obtains by taking into
account interactions between one magnetic moment and all its neighbors.)
Thus, we have $N^*=2N$ and consider the total potential energy
\begin{eqnarray}
U_i=\frac{\mu_i B^\prime}{N}\Sigma_{|i-j|=1}\left(1-\frac{5}{2}\alpha_{ij}^2\right)
+U_{g}(\theta_i),
\end{eqnarray}
where $\alpha_{ij}= N(\theta_i-\theta_j)/2\pi$. The Euler-Lagrange 
equations of motion are 
\begin{eqnarray}
\frac{M_TR}{N}\ddot{\theta}_i+\frac{\partial U_i}{R\partial\theta_i}=0.
\end{eqnarray}
This defines the system of equations for the vector
${\bf x}=(\theta_1,\theta_2,\cdots, \theta_N)$ given by
\begin{eqnarray}
\frac{M_TR}{N}\ddot{{\bf x}}+\frac{M_TM_H}{NR^2}{\bf x}=
\frac{5\mu B}{2N}\left(
\begin{array}{cccccc}
2 & -1 & 0 & \cdots & 0 & -1\\
-1&  2 & -1 & \cdots & 0& 0\\
 &  & \cdots\\
-1 & 0 & \cdots & 0 & -1 & 2
\end{array}
\right){\bf x}
\end{eqnarray}
The least stable eigenvector is ${\bf x}=(1,-1,1,\cdots,-1)$
(for $N$ even), for which the critical value of the magnetic field is 
\begin{eqnarray}
B^2_cM_H^2=\frac{1}{5}\left(\frac{M_T}{M_H}\right)\left(\frac{M_H}{R}\right)^4.
\label{EQN_BC2}
\end{eqnarray}
This condition is very similar to (\ref{EQN_BC}), and gives a commensurable
estimate
\begin{eqnarray}
\left.\frac{{\cal E}_B}{{\cal E}_k}\right|_c=\frac{1}{15}
\end{eqnarray}
and lifetime of rapid spin of the black hole.

\begin{figure}
\plotone{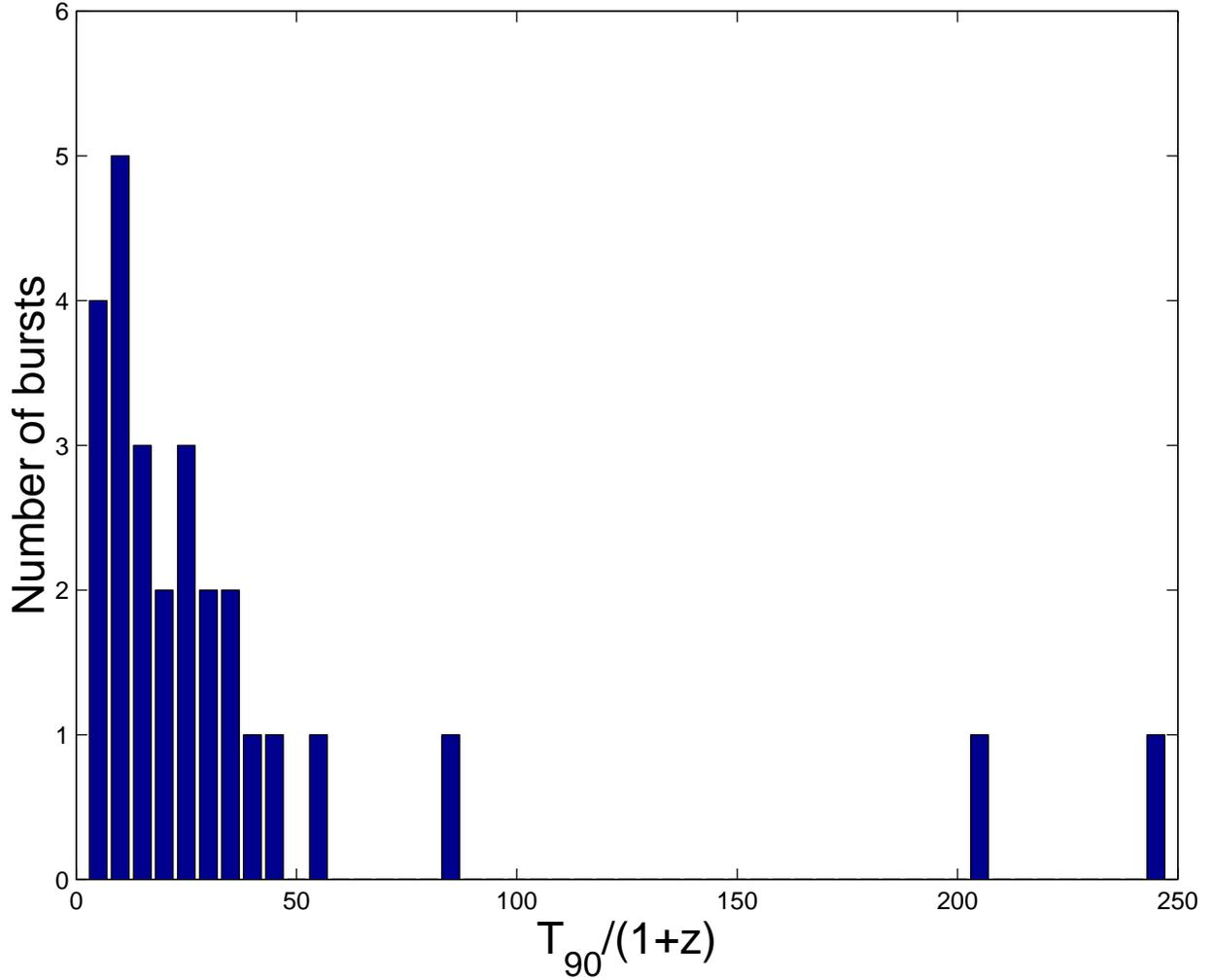}
\caption{Shown is the histogram of redshift corrected durations
of 27 long bursts with individually determined redshifts from their 
afterglow emissions. We identify these long durations with the
lifetime of rapid spin of a Kerr black hole in a state of
suspended accretion. These durations are effectively defined by the 
rate of dissipation of black hole-spin energy in the horizon, 
which is subject to a new magnetic stability criterion for the torus.
(Reprinted from M.H.P.M. van Putten, 2002, ApJ, 575, L71).}
\end{figure}

We propose to identify the redshift-corrected durations of long gamma-ray bursts 
and their contemporaneous emissions in gravitational radiation (Fig. 8)
with this new secular time-scale (\ref{EQN_LIFETIME}) of tens of seconds, set by 
the {stability-limited energy of the poloidal magnetic field supported
by the torus}.
 
A high-order approach can be envisioned, in which the inner and outer face 
of the torus are each partitioned by a ring of magnetized fluid elements.
This is of potential interest in studying instabilities in response to shear,
in view of the relative angular velocity $\Omega_+-\Omega_->0$. Magnetic
coupling between the two faces of the torus through aforementioned
tilt or buckling modes inevitably leads to transport of energy and
angular momentum from the inner face to the outer face of the torus
by the Rayleigh criterion. A detailed discussion of this instability
falls outside the scope of the present paper.

\section{Multipole mass moments in a torus in suspended accretion}
\label{sec:susaccrt}

As explained in \S \ref{sec:T-magnetosphere}, the torus is sandwiched between
a magnetic wall around the black hole and an outer torus magnetosphere. It is
hereby subject to competing torques on the inner and the outer face,
which introduces heating and, possibly, turbulent magnetohydrodynamical flow.
This is stimulated by a finite number of multipole mass moments produced by the
Papaloizou-Pringle instability in a torus of finite width.
The net rate of dissipation can be calculated in the suspended
accretion state, from balance of energy and angular momentum in combination
with emissions in gravitational radiation and magnetic winds. 

  A torus tends to develop instabilities in response to shear, which can be
  studied analytically in the approximation of incompressible fluid
  about an unperturbed angular velocity 
  \begin{eqnarray}
  \Omega=\Omega_a\left(\frac{a}{r}\right)^q,
  \end{eqnarray}
  where $q\ge 3/2$ denotes the rotation index and $a=(R_++R_-)/2$ denotes the
  major radius of the torus. In the inviscid limit, 
  irrotational modes in response to initially irrotational perturbations to the 
  underlying flow (vortical if $q\ne2$) shows the Papaloizou-Pringle instability
  \citep{pap84} to also operate in wide tori \citep{mvp02c}.
  The neutral stability curves of the resulting buckling modes are described
  by a critical rotation index $q_c=q_c(\delta,m)$ as a function of the
  slenderness parameter $\delta=b/2a$, 
  where $b=(R_{-}-R_{+})/2$ denotes the minor radius of the torus and
  $a=(R_-+R_+)/2$ the mean radius (Fig. 9), 
where $m$ denotes the azimuthal wave-number.

It will be appreciated that the torus is $m=0$ stable for perturbations of its
radius (in the mean). This is due to the frozen-in condition of magnetic flux-surfaces.
In terms of angular momentum transport, the horizon magnetic flux $\propto (M/a)^2$ 
hereby defines a black hole-to-torus coupling which is dominant over the coupling 
$\propto\Omega_T\simeq M/a^{3/2}$ of the torus to infinity. 

  Quadratic fits to the stability curves are
  \begin{eqnarray}
  q_c(\delta,m)=\left\{
  \begin{array}{rl}
  0.10(\delta/0.1)^2+1.73 & (m=2)\\
  0.26(\delta/0.1)^2+1.73 & (m=3)\\
  0.50(\delta/0.1)^2+1.73 & (m=4)\\
  0.80(\delta/0.1)^2+1.73 & (m=5)\\
  0.034m^2(\delta/0.1)^2+1.73     & (m>5)
  \end{array}\right.
  \label{EQN_FIT}
  \end{eqnarray}
  Instability sets in above these curves, stability below. For $m=2$, the critical 
  value $q_c=2$ obtains for $\delta=0.16,$
  associated with the Rayleigh stability criterion for the azimuthally symmetric 
  wave mode $m=0$. For large $m$, we use the numerical result of critical values
  $\delta=0.28/m$ for $q=2$.
\begin{figure}
%\plotone{p6a2.eps}
\plotone{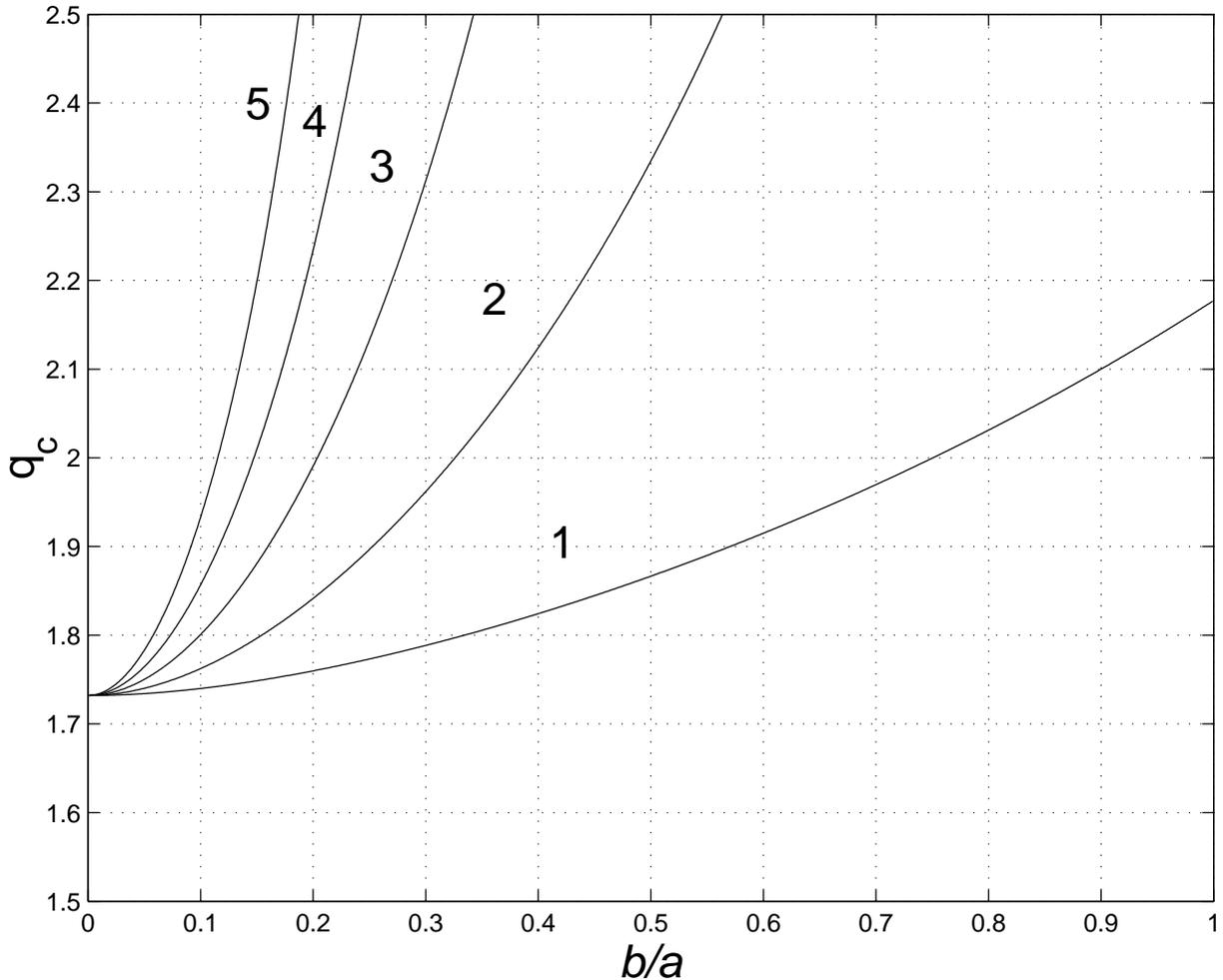}
\caption{Diagram showing the neutral stability curves for the non-axisymmetric
  buckling modes in a 
  torus of incompressible fluid, as an extension of the Papaloizou-Pringle 
  instabilities to finite slenderness ratios $\delta=b/2a>0$, where $b$ and $a$
  denote the minor and major radius of the torus, respectively.
  Curves of critical rotation index $q_c$ are labeled with azimuthal quantum 
  numbers $m=1,2,..$, where instability sets in above and stability sets in below.
  The central pressure produced by super- and subkeplerian
  motions of the inner and outer face of the torus is balanced by
  magnetic and thermal pressure at MeV temperatures. At finite
  slenderness $0<b/a<<1$, this produces
  instability to a finite number of Papaloizou-Pringle modes 
  (Reprinted from M.H.P.M. van Putten, ApJ, 575, L71).}
\end{figure}

  The central pressure of the torus is produces by differential rotation,
  due to a state of super-keplerian motion of the inner face and a state
  of sub-keplerian motion of the outer face. This pressure is balanced 
  by both magnetic and thermal pressure at MeV temperatures, which gives
  rise to finite slenderness $\delta>0$ and hence instability to a finite
  number of Papaloizou-Pringle modes.
  Thermal pressure alone gives the estimate (adapted from \cite{mvp02c})
  \begin{eqnarray}
  q\simeq 1.5 + 0.2
  %\times\left(\frac{a}{5M}\right)^3\left(\frac{M}{b}\right)^2
  \left(\frac{0.1}{\delta}\right)^2
  \left(\frac{kT}{2\mbox{MeV}}\right).
  \end{eqnarray}
  Magnetic pressure enhances this estimate to larger values of $q$.
  This shows that a torus of finite slenderness ($\delta<0.1$) 
  and a radius around $5M$ is unstable $(q>\sqrt{3})$ 
  to the formation of a finite number of multipole mass moments by
  the Papaloizou-Pringle instability. 
  {A torus with multiple mass moments potentially defines 
  a gravitational wave-spectrum consisting of several lines.}

\section{Energy emissions by the torus}
\label{sec:dissp}

The suspended accretion state in the case of a symmetric flux-distribution, 
given by equal fractions of
open magnetic flux on the inner and the {outer face,} gives rise to 
remarkably simple expressions for the predicted energy output in the
relevant channels.  To leading order the angular velocities of the inner
and outer faces are given in terms of the mean angular velocity of the
torus, $\Omega_T=(\Omega_{-}+\Omega_{+})/2$, and the slenderness ratio 
$\delta$, as $\Omega_{\pm}=\Omega_T(1\pm\delta)$.  Equation 
\ref{Lpm} implies that for a small slenderness ratio, 
$L_{-}\simeq \eta L_{+}$, where $\eta=\Omega_T/\Omega_H$ denotes the ratio 
of the angular velocity of the torus to that of the black hole.
In the limit of strong magnetohydrodynamical viscosity 
and small slenderness ratio we then have the asymptotic results \citep{mvp02d}
\begin{eqnarray}
E_{gw}/E_{rot}\sim \eta,~~~E_w/E_{rot}\sim\eta^2,~~~E_d/E_{rot}\sim\eta\delta,
\label{EQN_S4}
\end{eqnarray}
where $E_{gw}$, $E_{w}$, $E_{d}$ are defined below.

\mbox{}\\
{\em Gravitational radiation.} The major energy output from the torus is in
gravitational radiation,
\begin{eqnarray}
E_{gw}=6\times10^{53}\mbox{erg}\left(\frac{\eta}{0.1}\right)\left(\frac{M_H}{10M_\odot}\right).
\end{eqnarray}
A quadrupole buckling mode radiates gravitational waves at
close to twice the angular frequency of torus \citep{mvp02c}. 
These emissions
are relatively powerful, representing about 10\% of the rotational energy of
the black hole for the lifetime of the system. 
The associated mass inhomogeneity $\delta M_T$ in
the torus assumes a value commensurate with the inferred luminosity in gravitational
radiation. For quadrupole emissions, we have
%\begin{eqnarray}
$L_{GW}\simeq ({32}/{5})(M_H/R)^5(\delta M_T/M_H)^2,$
%\label{L_GW}
%\end{eqnarray}
where $\omega\simeq M_H^{1/2}R^{-3/2}$ denotes the orbital angular frequency
at a radius $R$, assuming approximately circular motion. The estimated gravitational
wave-luminosity is hereby produced by $\delta M_T\simeq 0.5\%M_H(R/5M_H)^{7/4}$. 
This corresponds to a mass inhomogeneity of 20\% for
a torus $M_T=0.2M_\odot$ around a black hole of mass $M_H=7M_\odot$.

\mbox{}\\
{\em Torus winds.} The energy output in torus winds is a factor $\eta$ less than
that in gravitational radiation, or
\begin{eqnarray} 
E_{w}=6\times10^{52}\mbox{erg}\left(\frac{\eta}{0.1}\right)^2\left(\frac{M}{10M_\odot}\right).
\end{eqnarray}
These powerful torus winds may produce hypernova remnants in the host environment,
e.g., a shell associated with a molecular cloud. They may be baryon loaded and deposit 
some torus matter onto the companion star as in the \cite{bro00} association of 
hypernovae to soft X-ray transients, and they may be important in collimating 
baryon poor jets produced by the black hole. Finally, these winds are potentially
relevant in r-processes \citep{lev93}. This suggests considering 
observational methods to determine $E_w$, from which to determine the system parameter
$\eta$, and hence the frequency of gravitational radiation. Of potential interest
are model dependent estimates of $E_w$ from their role in collimating outflows,
and calorimetry on hypernova remnants. The first can be pursued using existing
data on GRB beaming angles, which suggests a value of 
$\eta\simeq0.1$.
The second method
is potentially more reliable, but awaits further study on the identification and
observational aspects of hypernova remnants.

\mbox{}\\
{\em Torus temperature.} The energy output in thermal and neutrino emissions
is a factor $\delta$ less than that in gravitational radiation, or
\begin{eqnarray} 
E_{d}=10^{53}\mbox{erg}\left(\frac{\eta}{0.1}\right)
                               \left(\frac{\delta}{0.15}\right)
                               \left(\frac{M_H}{10M_\odot}\right).
\end{eqnarray}
Here, we refer to a fiducial value of $\eta\simeq0.1$ as before, as well as
a value $\delta\simeq0.15$ or less.
The latter is suggested by quadrupole radiation
of gravitational waves which requires $\delta\le 0.16$ 
at the threshold of Rayleigh stability, according to (\ref{EQN_FIT}).
This dissipation rate 
corresponds to a temperature of a few MeV (see eq. [\ref{Tav}] below), which
thereby produces baryonic winds. The torus winds considered here, therefore,
are baryon rich. 

\section{Pressure driven mass ejection from the torus}
\label{sec:massejection}
It has been  argued in \S \ref{sec:dissp} above that a fraction $E_d/E_{rot}$ of 
the black hole-spin energy is dissipated in the torus, thereby heating it to  
a temperature in excess of a few MeV.  This drives a powerful 
wind by the pressure gradients in the surface layers of the torus.
The resulting mass ejection opens magnetic 
field lines that pass through the outer Alfven point, and folds some of those in
the outer layers of the inner magnetosphere. This creates a change in poloidal
topology in the form of a coaxial structure of 
open flux tubes with opposite magnetic orientation (towards infinity).
Because the torus is rotating and magnetized, the ejection of the wind 
is partially anisotropic. Specifically, mass flux is generally
suppressed along magnetic field lines that are inclined toward the rotation axis,
and enhanced along field lines that are strongly inclined away from the axis
(Blandford \& Payne 1982; Romanova et al. 1997), owing to centrifugal forces.  
The details of the outflow depend on the heating and cooling rate of the corona,
and on its structure, and analysis thereof is beyond the scope of the present
discussion. In what follows, we provide a rough estimate for the mass flux expelled
in the vertical direction, speculate on the implications for opening of 
magnetic field lines, and discuss the consequences of mass ejection 
in the equatorial plane for the energy extraction process.

\subsection{An estimate of the vertical mass flux}

A torus having a temperature in excess of a few MeV, and an average 
density $\tilde{\rho}_b\sim 10^{11}(M_T/0.1 M_{\sun})/V_{21}$ 
gr cm$^{-3}$, where $V=10^{21}V_{21}$ cm$^3$ is the volume of the torus, 
cools predominantly by neutrino emission through electron and positron capture
on nucleons.  The cooling rate is given by (e.g., Bethe \& Wilson 1985)
\begin{equation}
\epsilon_{\rm cap}\simeq 10^{29}\tilde{\rho}_{b11}T_{10}^6\ \ {\rm ergs\ s^{-1}\ cm^{-3}},  
\label{eps}
\end{equation}
where $T_{10}$ is the average temperature in units of 10$^{10}$ K.  For a total energy
dissipation rate of $L=10^{52}L_{52}$ ergs, we then find an average temperature of
\begin{equation}
T_{10}\simeq 2 L_{52}^{1/6}(M_T/0.1M_{\sun})^{-1/6}.
\label{Tav}
\end{equation}

As will be shown below, the mass flux from the surface depends on the temperature 
profile in the neighborhood of the flow critical point, where the density is well 
below the average.
The latter can be determined in principle by equating the local heating and cooling 
rates, provided that adiabatic cooling there can be neglected (which we
find to be justified only if the temperature exceeds $\sim 2\times10^{10}$ K).  
While the fraction of the black hole spin-energy {dissipated} in 
the surface layers is not well constrained, a fraction of the energy of
neutrinos escaping from the dense regions -- well beneath the surface -- will be
deposited in the surface layers of the torus. The dominant absorption processes for
the latter are neutrino capture on neutrons and protons 
and pair neutrino annihilation into electron-positron pair. These processes
typically dominate at lower densities and higher temperatures.
In spherical
geometry, the heating rate is dominated by neutrino annihilation 
at densities below $\rho=6\times10^7(L_{52}/R_{\nu6}^2)^{1/2}(R_{\nu}/r)^6$ 
gr cm$^{-3}$, where $R_{\nu}=10^6R_{\nu6}$ cm is the radius of the 
neutrino production region \citep{lev93}. The cooling rate due to
electron-positron annihilation into 
neutrinos depends sensitively on temperature ($\epsilon_{\nu\bar{\nu}}\propto
T^9$), and dominates over the energy loss rate  by electron and positron capture on
nucleons at densities below $\rho=5\times10^6 
T_{10}^3$ gr cm$^{-3}$ \citep{lev93}.  Taking into account both 
processes, we find, up to a geometrical factor, that the critical point 
(see eq. [\ref{rho_c1}] below) will be maintained roughly at the average 
temperature given by eq. (\ref{Tav}). 

Now, the torus material should be a mixture of baryons and a light fluid (photons 
and electron-positron pairs in equilibrium).  The light and baryonic fluids will 
be tightly coupled, owing to the large Thomson depth.  Deep beneath its surface, 
the torus is in a hydrostatic equilibrium where the vertical gravitational force 
exerted on it by the black hole is supported by the baryon pressure $p_b=n_bkT$.
At baryon densities
\begin{equation}
\rho_b< \frac{p_lm_p}{kT}\simeq 10^{11} T_{10}^3\ \ \ {\rm gr\ cm^{-3}},
\end{equation}
the light fluid pressure,
\begin{equation}
p_{l}=2\times10^{25}T_{10}^4\ \ \ {\rm dyn\ cm^{-2}},
\end{equation}
exceeds the pressure contributed by the baryons.   At sufficiently 
shallow layers, the light fluid pressure gradient overcomes the vertical 
gravitational force, and the matter starts to accelerate.  The transonic 
flow should pass through a critical point.  To estimate 
the mass flux we calculate below the wind density and velocity at the critical
point. 

To simplify the analysis we neglect general relativistic effects.
Since we are merely interested in an estimate for the mass loss rate,
we need only to consider the properties of the flow in the neighborhood of the 
critical point.  Since the latter is located well within the light cylinder, 
we neglect rotation of 
the torus and the toroidal magnetic field. (This is not valid in general,
but is used only in regards to mass ejection from the upper and lower faces of the torus.)
Below, we find that the flow becomes 
mildly relativistic at the critical point. 
The MHD limit applies, in view of high electric conductivity in the
coupled light plus 
baryonic fluids, whereby 
the streamlines of the flow are along magnetic flux-surfaces.  Baryon
number conservation, viz., ${\bf\nabla}(n_b {\bf u}_p)=0$, where ${\bf u}_p$
denotes the poloidal 4-velocity ans $n_b$ the baryon number density, 
and Maxwell's equation, ${\bf\nabla}\times{\bf E}=0$, imply that the 
flux of baryons per unit magnetic flux is conserved:
\begin{equation}
(n_b u_p/B_p)^{\prime}=0.
\label{flx}
\end{equation}
Here $B_p$ is the poloidal magnetic field, and $^{\prime}$ denotes 
derivative along streamlines (i.e., ${\bf u}_p\cdot{\bf\nabla}$). 
In the limit of weak gravitational field, the projection of the momentum 
equation, $T^{i\nu}_{\ ;\nu}=0$, on the poloidal direction, and the use
of eq. (\ref{flx}) gives \citep{Cam86,takh90}
\begin{equation}
u_p(1-M^{-2})u_p^{\prime}=-\frac{\gamma^2}{1-a_s^2}[a_s^2(\ln B_p)^{\prime}+\psi^{\prime}],
\label{crit}
\end{equation}
where $\gamma^2= 1+u_p^2$ is the Lorentz factor of the flow, $\psi=GM_H/rc^2$ is
the gravitational potential,
\begin{equation}
a_s=\left(\frac{4 p_{l}}{12p_{l}+3\rho_bc^2}\right)^{1/2},
\label{a_s}
\end{equation}
is the sound speed of the mixed fluid (measured in units of $c$), 
and $M=u_p/c_s$, with $c_s=a_s/\sqrt{1-a_s^2}$ 
being the sound four speed, is the corresponding Mach number.
It is seen that the flow has a critical point at $M=1$.  We note that under the assumption 
made above, that the toroidal magnetic field can be ignored, this critical 
point coincides essentially with the fast magnetosonic point.  We now suppose that the flow 
passes through this critical point, which should be true in the case of a wind 
expelled along open magnetic filed lines. There the right-hand side of eq. (\ref{crit}) 
must also vanish. The latter condition can be solved to yield the sound speed at
the critical point:

\begin{equation}
a_{sc}^2=-\frac{\psi^{\prime}}{(\ln B_p)^{\prime}}=\frac{GM_H}{c^2r_c^2}
\frac{r^{\prime}}{(\ln B_p)^{\prime}},
\end{equation}
where $r_c\sim a$ is the radius of the critical point.  It is seen that the 
critical sound speed depends on the geometry of the field lines
in the vicinity of the critical point.  To obtain an order of magnitude estimate,
let us assume that $r^{\prime}\sim \xi_c/r_c$, and $(\ln B_p)^{\prime}\sim \xi_c^{-1}$,
where $\xi_c<r_c$ denotes the distance from the torus midplane to the critical point
along streamlines.  Then
\begin{equation}
a_{sc}\simeq \left(\frac{GM_H}{c^2r_c}\right)^{1/2}\left(\frac{\xi_c}{r_c}\right)= 
0.3\left(\frac{M_H}{10M_{\sun}}\right)^{1/2}r_{c7}^{-1/2}(\xi_c/r_c).
\label{a_sc}
\end{equation}
The critical density can be obtained now by employing eqs. 
(\ref{Tav}),  (\ref{a_s}) and (\ref{a_sc}):
\begin{equation}
\rho_{bc}=\frac{p_l}{c^2}\frac{(4-12a_{sc}^2)}{a_{sc}^2}
\simeq 1\times10^7\left(\frac{M_H}{10M_{\sun}}\right)^{-1}
\left(\frac{M_T}{0.1M_{\sun}}\right)^{-2/3}r_{c7}(\xi_c/r_c)^{-2}
L_{52}^{2/3}\ \ {\rm gr\ cm^{-3}}.
\label{rho_c1}
\end{equation}
The associated mass flux is given by
\begin{equation}
\rho_{bc}c_s\simeq1\times10^{17}\left(\frac{M_H}{10M_{\sun}}\right)^{-1/2}
\left(\frac{M_T}{0.1M_{\sun}}\right)^{-2/3}r_{c7}^{1/2}(\xi_c/r_c)^{-1}L_{52}^{2/3}
\ \ {\rm gr\ cm^{-2}\ s^{-1}},
\label{mflx}
\end{equation}
which corresponds to a mass loss rate of $\dot{M}\simeq 1\times10^{30}$ gr s$^{-1}$ for
a surface area of $A=0.1 a^2=10^{13}a_7^2$ cm$^2$.  
We conclude that during the $\sim 30$ s suspended accretion state the torus
will be partially evaporated. 

Finally, we note that the Alfven Mach number at the critical point is 
\begin{equation}
M_A=\left(\frac{16\pi p_l}{3B_p^2}\right)^{1/2}\simeq 0.07 (M_T/0.1M_{\sun})^{-1/3}
L_{52}^{1/3}B_{p15}^{-1}.
\label{M_A}
\end{equation}
We thus conclude that, for our choice of torus parameters, the outflow is sub-Alfvenic
(but not highly so) at the critical point.

\subsection{Creation of open field-lines}
\begin{figure}
%\plotone{f2002d}
%\plottwo{c_A}{c_B}
\plotone{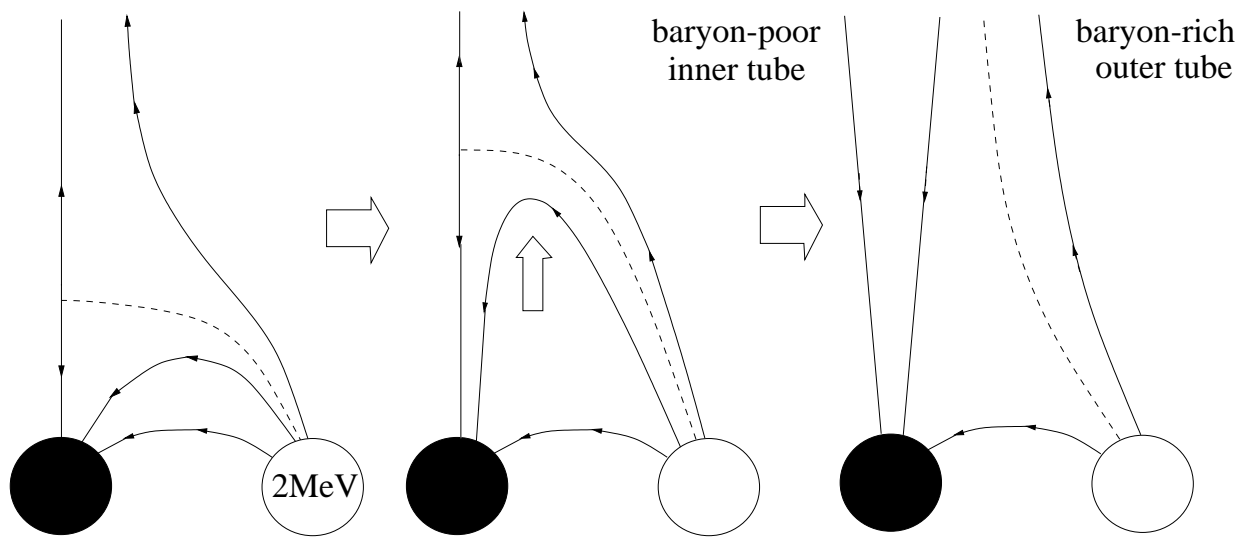}
\caption{The MeV temperature of the torus produces powerful torus winds. We
propose that these winds open an outer layer of flux-surfaces, supported  
by the inner face of the torus. This is schematically indicated by a stretch,
fold and cut of the separatrix (dashed lines) 
between the inner and outer flux-surfaces.
This leaves an open, inner tube of magnetic flux supported by the black hole
and an open, outer flux tube of magnetic flux supported by the inner face of
the torus. The inner tube is baryon-poor; the outer tube is baryon-rich.
The lower/upper sections of this structure contain parallel/antiparallel 
orientation of the poloidal magnetic field. The inner tube serves as an
artery for baryon-poor outflow from the black hole.}
% Its horizon half-opening angle
%$\theta_H$ defines the fraction of black hole-spin energy thus released.}
\end{figure}
The ejection of matter from the hot torus corona will result in the opening of
some magnetic field lines in the outer layers of the inner torus magnetosphere
(Fig. 10). By eq. (\ref{M_A}), plasma which streams along 
field lines that extend to large distances quickly reach the Alfven point.  
These field lines become nearly radial several scale height above the torus,
forming an open flux tube.  For field lines that converge toward the rotation axis 
of the black hole the above analysis is probably inapplicable, as the centrifugal 
force cannot be ignored.  Large pressure gradients in the corona would tend to 
push matter along
some of the magnetic field lines in the outer layers of the inner torus magnetosphere.
A combination of buoyancy and centrifugal forces may subsequently give rise to a twist 
of these field lines, some of which may ultimately fold and open, to form a region 
of oppositely directed magnetic field lines. Field lines thus created near the axis 
constitute the inner flux tube that extends from the horizon to infinity; those 
anchored to the torus now extend to infinity and constitute an outer flux tube. 
The inner and outer flux tubes have opposite magnetic orientation and are separated 
by a charge and current sheet, whenever the outer tube carries a (super-Alfv\'enic) 
wind to infinity.  Reconnection may occur in the boundary layer, which is of interest
in the rearrangement of the magnetosphere near the axis. 
This coaxial structure of open flux tubes is discussed in greater detail
in the next section.

A crucial issue is the opening angle of the inner flux tube on the horizon of the 
black hole, as it forms an artery for its spin-energy.  We consider
it probable that the opening angle depends only on the overall geometry of the
system (black hole mass and radius of the torus). A universal 
half-opening angle on the horizon gives rise to a standard fraction of black hole-spin
energy released through the inner flux-tube.

\subsection{Implications for energy extraction from the black hole}

The rate at which the torus catalyzes black hole spin-energy
has been calculated in \S \ref{sec:dissp} under the assumption that 
the torus magnetosphere
is force-free.  However, ejection of appreciable mass flux along magnetic 
field lines that thread the horizon may alter the extraction process.
Detailed study of ideal MHD flow in Kerr geometry is provided by Takahashi 
et al. (1990), and Hirotani et al. (1992).  They analyze the conditions
under which energy extraction occurs in a flow that starts with a zero 
poloidal velocity at the plasma source and is pulled inward by the black hole.  
They show that the MHD flow can carry a negative energy flux into the horizon
(which is equivalent to the condition that the energy given by eq. [\ref{E}] is
negative), provided that i) the angular velocity of magnetic
field lines lies in the range $0<\Omega_F<\Omega_H$, where $\Omega_H$ is the 
angular velocity of the black hole  (which is identical to the condition found
by \cite{bz77}), and ii) the Alfven point is located inside the ergosphere, which 
implies that inflow of negative electromagnetic energy exceeds 
inflow of positive kinetic energy.  The latter condition restricts the mass flux 
expelled along field lines connecting the torus and the horizon.  The vertical 
mass flux estimated above assumes maximum extraction efficiency.  We argue that 
inward mass ejection in the equatorial plane may be suppressed by the centrifugal 
barrier.  The details are complicated by virtue of general relativistic effects. 
Whether this is sufficient to allow energy extraction is not clear at present.
If not, it would mean that the rate at which the spin down energy of the hole is 
dissipated in the torus and, hence, its temperature, must be regulated by 
mass ejection.  

\section{Structure of the inner flux-tube}
\label{sec:opentubes}

In the preceding section we argued that mass ejection from the torus 
opens magnetic field lines on the outer layers of the inner torus magnetosphere.
This creates an open magnetic flux tube that extends from the horizon to infinity,
surrounded by an outer flux-tube that is anchored to the torus. The upper sections 
of this structure results
from a fold, stretch and cut in poloidal topology, whereby the inner and outer
tubes assume mutually antiparallel poloidal magnetic field. The lower section
remains unfolded, leaving parallel poloidal magnetic fields in the inner and 
outer flux-tubes (see Figs. 4, 10). This additional structure was not analyzed
in \citep{mvp02}.  In what follows, we describe the structure of 
the inner tube in some detail. 
 
\subsection{Asymptotic boundary conditions on the horizon and at infinity}

The inner flux tube satisfies slip/slip boundary conditions both on the 
horizon and at infinity.  It is well known that in the limit of infinite 
conductivity, every magnetic surface must rotate rigidly (although the 
angular velocity of different flux surfaces should not be the same). Finite 
resistivity effects, however, may give rise to a differential rotation along 
magnetic flux tubes
(implying ${\bf E}\cdot{\bf B}\neq0$).  To make our analysis general, we shall
allow for such a differential rotation, and denote the angular velocities of 
a given flux surface on the horizon and at infinity by  $\Omega_{F+}$ and $\Omega_{F-}$,
respectively.   

As stated above, 
we anticipate the horizon half-opening angle $\theta_H$ of the inner flux tube 
to be sufficiently
small so that the small-angle approximation applies in solving Maxwell's equations.
From equation (\ref{Max-r}) we obtain
\begin{equation}
2\pi\int{j^r\sqrt{-g}d\theta} =\frac{\Delta}{2 \rho^2}\sin\theta F_{r\theta}
\equiv \frac{1}{2}B_T.
\label{I}
\end{equation}
It can be readily shown (e.g., \cite{bz77}) that in 
the force-free limit, viz., $F_{\mu\nu}j^{\nu}=0$, 
the Boyer-Lindquist toroidal magnetic field, $B_T$, is conserved along magnetic flux 
surfaces $\Psi(r,\theta)$, and that the current contained within a magnetic flux surface 
is given exactly by eq. (\ref{I}), viz., $I(\Psi)=(1/2)B_T$ and, hence, is 
also conserved.  Current conservation along streamlines is not
guaranteed in general, however,  even in the limit of infinite conductivity, 
since inertial effects may give rise to cross field currents.  At any rate, 
beyond the fast magnetosonic point of the inflow (outflow), the poloidal 
current becomes radial asymptotically, as it is carried purely by the inflowing 
(outflowing) Goldreich-Julian charges (see below).  By employing the 
asymptotically frozen-in condition on the horizon, $F_{r\theta}
=(\Sigma/\Delta)(\Omega_H-\Omega_{F+})F_{\phi\theta}$ 
(see eq. [\ref{ratio1}] with $\beta=-\Omega_H$), we 
obtain the poloidal current flowing through the horizon:  
\begin{equation}
I_H=\frac{(r_H^2+a^2)}{2(r_H^2+a^2\cos^2\theta)}
\sin\theta F_{\phi\theta}(\Omega_H-\Omega_{F+}),
\label{IH}
\end{equation}
Assuming the ZAMO radial magnetic field, 
$B_r=-F_{\phi\theta}/(\Sigma\sin\theta)$, to be independent 
of $\theta$ near the axis, we can calculate the total magnetic flux through 
the horizon in one hemisphere: $\Psi=2\pi A_{\phi}
=\int{B_r\sqrt{g_{\phi\phi}g_{\theta\theta}}d\theta d\phi} 
\simeq \pi\Sigma\sin^2\theta B_r = -\pi\sin\theta F_{\phi\theta}$.    
Eq. (\ref{IH}) then yields to leading order, 
\begin{equation}
I_H\simeq -(\Omega_H-\Omega_{F+}) A_{\phi}.
\label{IH-app}
\end{equation}
Likewise, the frozen-in condition at infinity gives 
$F_{r\theta}=\Omega_{F-} F_{\phi\theta}$, leading in the small 
angle approximation to,
\begin{equation}
I_{\infty}= (1/2)\sin\theta F_{\phi\theta}\Omega_{F-}\simeq-\Omega_{F-} A_{\phi}.
\label{Iinf-app}
\end{equation}
%where $\Omega_-$ denotes the angular velocity of the upper section which
%approaches infinity.
The electric charge distribution in the inner flux tube can be obtained from 
Maxwell's equation: $F^{t\mu}_{\ \ ;\mu}=4\pi j^{t}$.  At small angles this 
equation is given, to a good approximation, by eq. (\ref{Max-t}) in the Appendix 
(cf. Van Putten 2001).  Assuming as before that the ZAMO radial magnetic field 
is independent of $\theta$ near the 
axis, we obtain from eq. (\ref{Max-t}) the Goldreich-Julian charge density:
\begin{equation}
\rho_e=\alpha^2 j^t=-\frac{(\Omega_F+\beta)B_r\cos\theta}{2\pi},
\label{rho_e}
\end{equation}
where $\Omega_F$ denotes the local angular velocity of the flux-surface at hand.
Evidently, the electric charge changes along the inner flux tube from 
$\rho_{eH}=(\Omega_H-\Omega_{F+})B_r/2\pi$ on the horizon, to 
$\rho_{e\infty}=-\Omega_{F-} B_r/2\pi$ at infinity, and vanish at the radius
at which $-\beta=\Omega_F$, corresponding to a locally zero
angular momentum state of the flux-surface. 
This is illustrated in fig. 4.  Using eq.  (\ref{Max-r}) 
and (\ref{rho_e}), one finds $j^r= j^t v^r$, 
where $v^r=1$ at infinity, and $v^r= -\alpha^2$ on the horizon (see Appendix).
{\em This implies that the current on the horizon and at infinity is purely due  
to convection of Goldreich-Julian charges.}  A similar conclusion was drawn by 
Punsly and Coroniti (1990).

\subsection{Two steady state limits}

Current closure of $I_\infty$ through the inner tube over the outer tube with 
equal magnetic flux of opposite sign gives rise
to a differentially rotating inner tube, described by
\begin{eqnarray}
\Omega_{F+}-\Omega_{F-}=\Omega_H-2\Omega_T
\label{EQN_OMa}
\end{eqnarray}
\citep{mvp02}. This assumes the force-free limit, whereby current is
conserved along flux-surfaces and $I_H=I_\infty$. 

Alternatively, current closure of $I_\infty$ through the inner tube over the outer 
tube and the outer torus magnetosphere allows for the limit of a uniformly rotating
inner tube. This corresponds to the limit of infinity conductivity.
%$\Omega_F=\Omega_F(A_\phi)$ on each flux-surface.
Combined with the force-free limit, this approach was used by 
\cite{bz77} and Phinney (1983) to construct their force-free solutions and determine
the efficiency of the energy extraction process.  Near the axis 
Eqs. (\ref{IH-app}) and (\ref{Iinf-app}) yield to leading order,
\begin{eqnarray}
\Omega_{F+}=\Omega_{F-}=\Omega_H/2.
\label{EQN_OM}
\end{eqnarray}
It will be appreciated that this implies nearly maximal energy extraction
rate on the inner tube. 

Punsly \& Coroniti (PC; 1990) argue that the assumption made by 
\cite{bz77} and Phinney (1983), namely that the magnetosphere is 
force-free in the entire region between the horizon and infinity, violates the 
principle of MHD causality.  Their argument relies on the observation that the 
inflow must become superfast on the horizon and, therefore, cannot communicate 
with the plasma source region (e.g., the gap in the Blandford-Znajek model).  
They concluded that the use of the Znajek frozen-in condition on the horizon 
to determine $\Omega_F$ is unphysical, and that $\Omega_F$ must be determined 
by the dissipative process that leads to ejection of plasma on magnetic field 
lines. This point is of interest, as the black hole has a finite, though
secular lifetime of rapid spin as an inner engine to gamma-ray bursts. 
In what follows we reexamine this issue. 

We remark that MHD causality prohibits infinitely rigid structures (the Alfv\'en 
velocity is bounded by the velocity of light).
% whereby (\ref{EQN_OM}) holds only
%approximately. 
This is true in particular over distances much larger than
the system size. Nevertheless, we may examine a steady-state force-free limit
$\Omega_{F+}\simeq\Omega_{F-}\simeq\Omega_H/2$ as a close approximation 
on scales comparable to the system size, modified only by a differentially
{rotating gap which injects} the associated electric current in
a region where frame-dragging $\beta$ is pronounced.

The current flow in the inner tube is created in a slightly differentially
rotating gap in a neighborhood of $\Omega_F=-\beta$ between two Alfv\'en 
surfaces. The gap size (and hence current output) 
is determined by the local degree of differential frame-dragging in $\beta$.
As the magnetosphere around the black hole is transparent to $\beta$, the
gap remains in causal contact with the angular velocity of the black hole:
a change in the asymptotic value $-\beta=\Omega_H$ on the horizon is 
associated with a change in $\beta$ throughout the surrounding spacetime.
This indicates that the gap is subject to changes $\delta\Omega_H$ under 
general conditions.

%If $\Omega_{F+}\simeq\Omega_{F-}\simeq\Omega_H/2$ is 
If $I_H\ne I_{\infty}$, the inner 
tube develops a response by charge conservation: time-variable or in
steady state, depending on the absence or presence of cross-field currents 
(i.e., $j_{\theta}$) between the inner and the outer tube. 
Conceivably, the plasma injection process maintains the required 
cross-field currents, and there exist multiple current closure paths.
In view of (\ref{IH-app}) and (\ref{Iinf-app}), note that
this implies additional
differential rotation in the inner tube with accompanying dissipation 
processes.
In the absence of cross-currents, there results a time-variable adjustment 
of the two Alfv\'en surfaces which delimit the gap by application of Gauss' 
theorem to, respectively, the black hole plus lower section and the upper 
section. This causes adjustment of the current injected into the lower and 
the upper sections. In steady state, this recovers $I_H=I_{\infty}$. 
This argument leaves open the possibility that the gap is time-variable, 
however, especially on the light crossing time-scale.

\subsection{Output power through the inner tube}

The inner tube releases black hole-spin energy at a certain rate, set by
the horizon half-opening angle $\theta_H$ shown in Figs. 4 and 10.
Forementioned force-free limits (\ref{EQN_OMa}-\ref{EQN_OM}) define 
bounds for the bipolar outflows of black hole-spin energy through the
inner tube, i.e.,
\begin{eqnarray}
\Omega_{F+}(\Omega_H-\Omega_{F+})A_{in}^2\le S_m\le\frac{1}{4}\Omega_H^2A_{in}^2,
%S_m\le\frac{1}{4}\Omega_H^2A_{in}^2,
\label{EQN_PF}
\end{eqnarray}
where $2\pi A_{in}$ is the net magnetic flux on the horizon associated with 
the inner tube. Hence, we have a fraction 
\begin{eqnarray}
\frac{E_j}{E_{rot}}\simeq\frac{1}{4}\theta_H^4
\label{EQN_small_1}
\end{eqnarray}
of the rotational of the black hole (see eqs. [\ref{torque}] and [\ref{Lpm}]).

The observed true GRB energies cluster around $3\times 10^{50}$erg
\citep{fra01}.
This corresponds to $E_j=2\times 10^{51}$erg $\times (0.16/\epsilon)$, where 
$\epsilon$ denotes the efficiency of conversion of
kinetic energy-to-gamma rays. In our model of GRBs from rotating black holes,
this observational constraint introduces the small parameter
\begin{eqnarray}
\frac{E_j}{E_{rot}}\simeq 10^{-3}\left(\frac{7M_\odot}{M}\right)
\left(\frac{0.16}{\epsilon}\right).
\label{EQN_small_2}
\end{eqnarray}
This corresponds to a horizon half-opening angle $\theta_H\ge 15^o$;
$\theta_H\simeq35^o$ associated with the output from the gap in
(\ref{EQN_OMa}).

We propose to identify $\theta_H$ with the curvature in poloidal topology
of the inner torus magnetosphere \citep{mvp02d}, $\theta_H\sim M/R$
within a factor close to one. In this event, we have
\begin{eqnarray}
\frac{E_j}{E_{rot}}\sim \frac{1}{4}\left(\frac{M}{R}\right)^4.
\end{eqnarray}
A spread in torus radius by a factor of about two corresponds to a
spread in energies in baryon-poor outflows by about one order of magnitude, 
consistent with the observed spread in true GRB energies. 

The fate of the Poynting flux-outflow is determined by specific mechanisms
for dissipation, as these may arise downstream of the upper section of the
open flux-tube.

\section{Structure of the outer tube}
\label{Sec:outertube}
The open field lines that are anchored to the torus form the outer flux tube.  
The baryon-rich material ejected from the torus, derived from the spin-energy
of the black hole (see \S \ref{sec:massejection}), flows along those open 
field lines to infinity.
We have estimated this to constitute a substantial mass loss. In the ideal MHD 
approximation the angular velocity of the outer flux-tube is conserved along 
magnetic field lines (on the scale of the system) and, therefore, equals that of the 
torus, viz.,  $\Omega_F=\Omega_T$.  
Above the Alfven point the ratio of toroidal and poloidal field in the outflow
is (see eq [\ref{eq:Bphi/Br}] with $\Omega_F=\Omega_T$ and $v_{\phi}\rightarrow r^{-1}$)

\begin{equation}
\frac{F_{r\theta}}{F_{\phi\theta}}=\frac{\Omega_T}{v_r}.
\end{equation}
As argued above, the direction of the poloidal magnetic field in the outer 
flux tube is opposite to its direction in the inner tube.  Thus, if the 
torus and the black hole rotate in the same direction, as envisioned here, 
then the toroidal magnetic field $F_{r\theta}$ have opposite 
orientations in the inner and outer tubes.  
The Goldreich-Julian charge density in the outer tube is 
\begin{equation}
\rho_e=\frac{\Omega_T B_r\cos\theta}{2\pi},
\label{rho_ee}
\end{equation}
and is opposite in sign to the outflowing charges in the inner flux tube.  These charges
carry the current flowing along the open field lines from the torus.  
This region of the torus magnetosphere is equivalent to a pulsar magnetosphere.

We conclude that the outer tube forms out of the outer layers of the inner
torus magnetosphere (the magnetic wall around the black hole), 
and joins the open field-lines in 
the outer torus magnetosphere in creating a super-Alfv\'enic baryon-rich
wind to infinity. It conceivably contributes to collimation of the baryon-poor
outflows in the inner flux-tube.

\section{The interface between the inner and the outer flux-tube}
\label{Sec:interface}
The creation of the open magnetic flux tubes from the closed torus magnetosphere 
topologically represents folding of magnetic field lines in the upper section of
the inner/outer flux-tube, which gives rise to an antiparallel orientation of the
poloidal magnetic field. This is accompanied by a cylindrical
current and charge sheet that accounts for the jump in the electric and 
magnetic fields across the interface.
The lower section of the inner/outer flux-tube which connects to the horizon has,
instead, a parallel orientation between the poloidal magnetic field.
In the perfect MHD limit, the properties of the interface are described by jump
conditions as follow from Maxwell's equations, $F^{t\mu}_{\ \ ;\mu}=4\pi j^{t}$, i.e.:
\begin{equation}
4\pi\sigma_e=[\Omega r\sin\theta B_r]= r\sin\theta\Omega_T B_{r+}
-r\sin\theta(\Omega_H/2) B_{r-},
\end{equation}
where $B_{r+}$ ($B_{r-}$) denotes the radial magnetic field near the interface 
in the outer (inner) flux tube.  The poloidal and 
toroidal interface currents are, likewise,
\begin{eqnarray}
\begin{array}{rl}
4\pi J^r&=[B_\phi]=\left[\frac{\Omega r\sin\theta B_r}{v^r}\right],\\
4\pi J^{\phi}&=-[B_r].
\end{array}
\label{EQN_IP}
\end{eqnarray}
The poloidal current (\ref{EQN_IP}) results beyond the
Alfv\'en point, where the wind transports angular
momentum outwards to infinity.
The ouflow in the inner tube is expected to be relativistic ($v^r=1$).  
If the baryon rich outflow from the torus also becomes relativistic, then 
the latter equation imply that $J^r=\sigma_e$, namely the poloidal current 
in the boundary layer is solely due to the outflowing surface charges. 
The poloidal current sheet -- possibly further carrying a poloidal current -- 
is a potential site for reconnection of magnetic field lines,
which would convert magnetic enegy in the inner flux tube into kinetic energy.  

\subsection{Dissipation in the folded upper section of the flux-tubes}

The upper interface between the inner and the outer flux tubes is subject to 
antiparallel magnetic fields and strong differential 
rotation (whenever $\Omega_T\ne\Omega_H/2$. It is therefore a 
potential site for reconnection of magnetic field lines,
and may give rise to conversion of a fraction of the magnetic enegy
in the inner flux tube into kinetic energy.  In order to asses the 
fraction of magnetic energy which is converted in the interface, a detailed 
reconnection model is required.  Here, we point out that the reconnection 
time is limited by the crossing time af an Alfven wave across the inner
tube, which is typically very short.  If this limit applies then an appreciable 
fraction of magnetic energy will be dissipated.  (The reconnection rate   
could be much smaller.) In any case, the power
extracted from the hole by the inner tube is uncertain, because the angular
velocity of the inner tube on the horizon is unknown.  
In or near the lowest energy state of the gap, the inner tube mediates a power 
$S_m$, as given by eq. (\ref{EQN_PF}).

The electric fields produced by the reconnection process in the interface
may inject plasma, including baryons that should be present 
in the boundary layer, into the inner tube.  This would lead to mass 
loading of the inner tube and additional conversion of the Poynting flux.  The amount 
of baryonic contamination by this process is yet an open issue. 

Additional dissipation may arise from a differential rotation of the inner
flux tube, as discussed in \citep{mvp02}.  It is clear
that if the system fluctuates over a time scale $\Delta t$, then nonlinear 
disturbances may induce differential 
rotation over length scales $< c \Delta t$ (which corresponds
to $10^4$ gravitational radii for the entire life time
of the system).  Such electromagnetic 
fronts should be highly dissipative by virtue of the large parallel 
electric fields associated with the differential rotation of the magnetic
flux tubes.  It is anticipated that, under the conditions envisioned here,
copious pair and photon productions would ensue inside the 
differentially rotating fronts. 

\section{Observational consequences}
\label{Sec:conclusion}

We have described in some detail the topology, lifetime and emissions
of black hole-torus sytems in the suspended accretion state.
Fig. 11 summarizes the energy transport and conversion of black hole-spin energy, 
catalyzed by the torus. Most energy is dissipated in the horizon, while the major
output is converted by the torus into gravitational radiation, winds and 
neutrino emissions (as well as thermal emissions). A minor output is released in
baryon-poor outflows as input to the observed GRB-afterglows. 
The analysis is based largely on equivalence to pulsar magnetospheres.
The suspended accretion state develops against a magnetic wall around rapidly 
rotating black holes. This is based on a uniform magnetization of the torus, 
represented by two oppositely oriented current rings in a torus formed in
black hole-neutron star coalescence and core-collapse hypernovae.

\begin{figure}
%\plotone{f2002d}
%\plottwo{c_A}{c_B}
\plotone{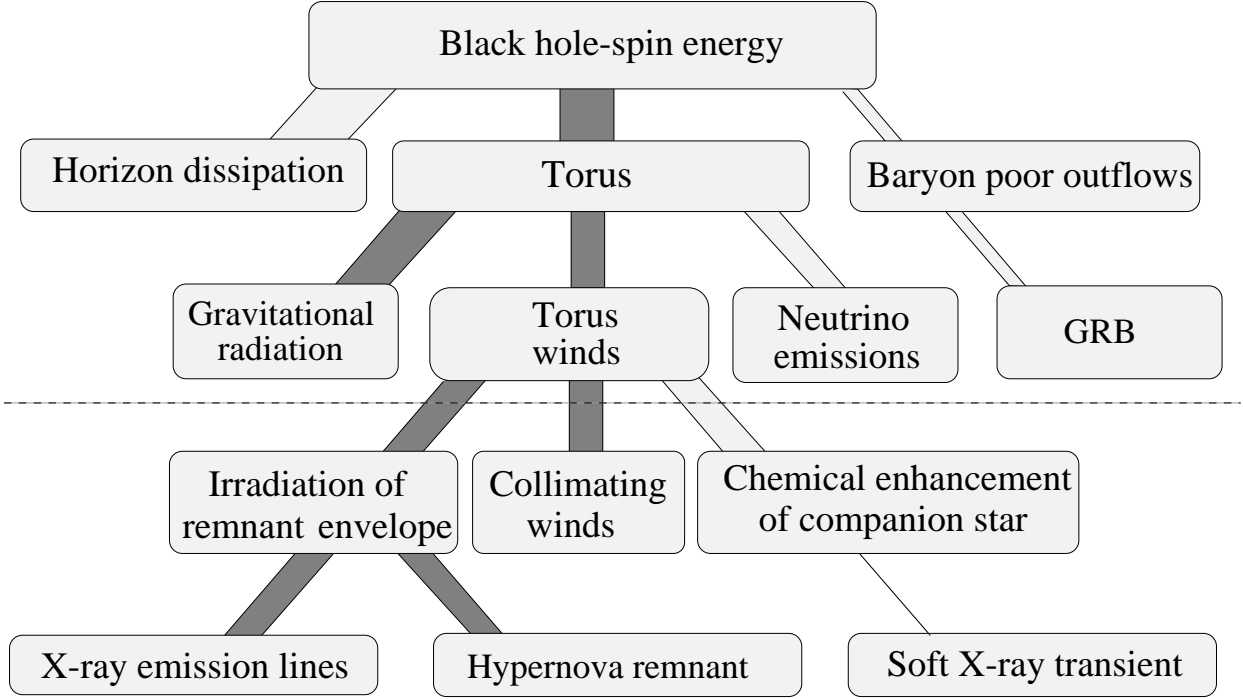}
\caption{Overview of energy transport and conversion of black hole-spin
energy catalyzed by a uniformly magnetized torus. Most of the spin-energy 
is dissipated in the horizon -- an unobservable sink of energy -- at a rate
which is limited by a new magnetic stability criterion for the torus. This
process effectively defines the lifetime of rapid spin of the black hole.
Most of the spin-energy released is incident on the inner face of the torus, 
while a minor 
fraction forms baryon-poor outflows through the inner flux-tube to infinity.
We associate the latter with the input to the observed GRB-afterglow emissions. 
The torus converts its input primarily into gravitational radiation and, to a 
lesser degree, into winds, thermal and neutrino emissions. Direct measurement of 
the energy and frequency emitted in gravitational radiation by the upcoming
gravitational wave-experiments provides a calorimetric compactness test for
Kerr black holes (dark connections). Channels for calorimetry on the torus 
winds are indicated below the dashed line, which are incomplete or 
unknown. They provide in principle a method for constraining the angular 
velocity of the torus and its frequency of gravitational radiation. This is 
exemplified by tracing back between torus winds and their remnants (dark 
connections). As the remnant envelope expands, it reaches optical depth of 
unity and releases the accumulated radiation from within. This continuum 
emission may account for the excitation of X-ray line emissions seen in GRB 
011211, which indicates a torus wind energy of a few times $10^{52}$erg. 
Matter ejecta ultimately leave remnants in the host molecular cloud, 
which remain to be identified. Torus winds may further deposit a fraction of 
their mass onto the companion star \citep{bro00}, 
thereby providing a chemical enrichment
in a remnant soft X-ray transient.}
\end{figure}

The proposed emissions in gravitational radiation are powered by the spin-energy
of the black hole, which renders these emissions a powerful new source for the
upcoming gravitational wave-experiments. This output is contemporaneous with 
GRBs from baryon poor outflows, and hence have the same durations as the 
redshift corrected durations of long GRBs -- tens of seconds, due to the
condition of magnetic stability of the torus.

Calorimetry on the emissions in gravitational radiation provides a rigorous 
compactness test for Kerr black holes \citep{mvp01b}, which can be pursued by
upcoming gravitational wave-experiments.
We emphasize that our predictions on the energies and duration of
emissions in gravitational radiation are robust and are independent of an 
association with GRBs. Hypernovae could be over-abundant as burst sources of 
gravitational radiation, with only a small fraction making successful GRBs.
Calorimetry on wind energies provides a method for constraining the angular
velocity of the torus and, hence, its quadrupole emissions in gravitational
radiation. Possible channels are recently observed X-ray line emissions
in GRB 011211 \citep{ree02}, hypernova remnants -- which remain to be
identified -- and by association with soft X-ray transients with chemically 
enhanced companion stars \citep{bro00}.
 
X-ray and radio shells that may result from the interaction of the baryon rich 
torus winds with ambient matter, as well as the gamma-ray 
emission expected to be produced by the baryon poor outflows in the inner tube,
should be sought for \citep{lev02}.

Below we summarize our findings and draw additional conclusions:
\mbox{}\\

1. {Most of the black hole-luminosity is incident on
the torus by topological equivalence to pulsar magnetospheres, when the
black hole spins rapidly.}
We identify a new magnetic instability, which defines a minimum
lifetime $T$ of tens of seconds of the spin of the black hole,
corresponding to a large system parameter $T/M\sim 10^6$. 

2. At MeV temperatures of the torus, a torus with a minor radius of less than
about the linear size of the black hole becomes unstable to the formation of a
finite multipole mass moments by the Papaloizou-Pringle instability and, 
conceivably, other modes.
The torus hereby converts a major fraction $E_{gw}/E_{rot}\simeq 10\%$ into 
gravitational radiation, {conceivably at multiple lines.} 
At a source distance of 100Mpc, these emissions over 
$N=2\times 10^4$ periods give rise to a characteristic
strain amplitude $\sqrt{N}h_{char}\simeq 6\times 10^{-21}$ \citep{cow02}.

3. A minor fraction $E_j/E_{rot}\sim 10^{-3}$ of black hole-spin energy is 
released in baryon poor outflows. We attribute this fraction to curvature in 
poloidal toplogy, whereby
$E_j/E_{rot}\simeq (1/4)(M/R)^4$ is standard for a universal 
horizon half-opening angle $\theta_H\simeq M/R$ of the associated open magnetic 
flux-tube. These outflows
are probably not high-sigma -- the ratio of Poynting flux-to-kinetic energy
flux -- owing, {in part,} 
to magnetic reconnection in an interface with
baryon-rich winds flowing along the surrounding outer flux-tube.

4. Causality in the process of spin-up of the torus by the black hole follows by
topological equivalence to pulsar magnetospheres \citep{mvp99}.
In response, the black hole spins down by conservation of energy and angular momentum.
This establishes causality in the extraction of black hole-spin energy by 
horizon Maxwell stresses as proposed by \cite{bz77}. 
Causality in the formation of baryon poor outflows from
the open magnetic flux-tube is due to transparency of the magnetosphere to 
the gravitational field: {the associated current injection
is subject to current continuity between asymptotic boundary conditions on the horizon and 
at infinity by Gauss' theorem, and
regulated by differential frame-dragging and current closure at infinity.}
%%Transparency of the magnetosphere to frame-dragging ensures that
%%the angular velocity of the inner tube 
%($\simeq\Omega_T$ to $\Omega_H/2$ on the 
%linear scale of the system) 
%%is causally coupled to the angular velocity of the black hole under general conditions.}

5. Calorimetry on the predicted energy output in gravitational radiation by the
upcoming gravitational wave experiments provides a method for identifying
Kerr black holes as objects in nature when $2\pi\int_0^{E_{gw}}f_{gw}dE>0.005$
\citep{mvp01b}, where $f_{gw}$ denotes the quadrupole frequency of gravitational
waves. Current experiments consist of
layer {interferometric detectors LIGO, VIRGO, TAMA and GEO, bar and sphere detectors.}
An individual source is band limited in gravitational radiation in terms of
a frequency sweep of about 10\%, corresponding to {emission of the first
50\% of the output in gravitational radiation from a maximally spinning Kerr black 
hole. Collectively, black hole-torus systems are conceivably luminous in
multiple frequencies with broad distributions owing to a spread in black hole mass.}

6. Frequencies of quadrupole gravitational
radiation $f_{gw}\simeq 470\mbox{Hz}$ 
$(E_w/4\times 10^{52}\mbox{erg})^{1/2}$ $(7M_\odot/M_H)^{3/2}$
can be predicted by calorimetry on the torus wind energies $E_w$.
Calorimetry on X-ray line emissions point towards frequencies around
500Hz \citep{mvp02d}. 

7. Calorimetry on $E_w$ may also be pursued by 
calorimetry on hypernova remnants, given the observed supernova
association (e.g., \cite{blo02}).
Our model suggests the ejection of the remnant stellar envelope 
as a shell with kinetic energy
$0.5\beta E_w\sim 3\times 10^{51}$erg$(\beta/0.1)(M_H/10M_\odot)(\eta/0.1)$.
Here, $\beta$ denotes the initial radial velocity $\beta$ relative to the velocity
of light, in response to the impact of $E_w$ from within.
For a supernova association to molecular clouds, 
see, e.g., 
\cite{chu90,wan91}; a hypernova or gamma-ray burst association to 
molecular clouds has been considered by
\cite{efr98,wan99,dun01,lai01,che02,pri02}. 
Determining kinetic energies of these remnants
can be pursued analogously to studying supernova
remnants, in the radio and X-ray. Chandra-X observations may hereby identify
hypernova remnants in X-ray bright (super)shells around
a black hole-binary, possibly in the form of a soft X-ray transient. 
%This is conceivably examplified by the X-ray binary in the remnant RX J050736-6847.8
%observed by ROSAT \citep{chu00}.

{\bf Acknowledgement.} The first author acknowledges constructive 
comments from G. Mendell and B.-C. Koo.
This research is supported by NASA Grant 5-7012, 
a NATO Collaborative Linkage Grant and an MIT C.E. Reed Fund. 
The LIGO Observatories were constructed by the California Institue of Technology
and Massachusetts Institute of Technology with funding from the National Science
Foundation under cooperative agreement PHY 9210038. The LIGO Laboratory operares
under cooperative agreement PHY-0107417. This paper has been assigned LIGO
Document Number {LIGO-P020030-00-R.}

\break

\appendix
\section{ Ideal MHD in Kerr geometry}
We express the Kerr metric in Boyer-Lindquest coordinates with the following notation:
\begin{equation}
ds^2= -\alpha^2dt^2 + \tilde{\omega}^2(d\phi+\beta dt)^2 + \frac{\rho^2}{\Delta}dr^2 + 
\rho^2 d\theta^2,
\label{metric}
\end{equation}
where $\alpha=\rho\sqrt{\Delta}/\Sigma$ is the lapse function, 
$\tilde{\omega}^2=(\Sigma^2/\rho^2)\sin^2\theta$,
$-\beta=2aMr/\Sigma^2$ is the angular velocity of a ZAMO with respect to a distant 
observer , with $\Delta=r^2+a^2-2Mr$, $\rho^2=r^2+a^2\cos^2\theta$, and 
$\Sigma^2=(r^2+a^2)^2-a^2\Delta\sin^2\theta$.  The parameters $M$ and $a$ are the mass
and angular momentum per unit mass of the black hole.

We denote by $n$, $p$, $\rho$, $h=(\rho+p)/n$, the proper particle 
density, pressure, energy density, and specific enthalpy, respectively.  
The stress-energy tensor then takes the form:
\begin{equation}
T^{\alpha\beta} = hnu^{\alpha}u^{\beta} - pg^{\alpha\beta}+\frac{1}{4\pi}
(F^{\alpha\sigma}F^{\beta}_{\sigma}+\frac{1}{4}g^{\alpha\beta}F^2),
\end{equation}
where $u^{\alpha}$ is the four-velocity, and $F_{\mu\nu}$ is the electromagnetic 
tensor which satisfies Maxwell equations.  The dynamics of the MHD flow is then 
governed by the equation 
\begin{equation}
T^{\mu\nu}_{\ \ ;\nu}=0,
\label{Tmn}
\end{equation}
subject to conservation of particle flux,
\begin{equation}
(nu^{\alpha})_{;\alpha}=\frac{1}{\sqrt{-g}}(\sqrt{-g}nu^{\alpha})_{,\alpha}=0.
\label{continuity}
\end{equation}
The ideal MHD assumption (i.e., infinite conductivity) imposes the additional constraint,
\begin{equation}
u^{\alpha}F_{\alpha\beta}=0.
\label{ideal}
\end{equation}
The stationary axisymmetric flow considered here is characterized by two Killing vectors:
$\xi^{\mu}=\partial_t$ and $\chi^{\mu}=\partial_{\phi}$.  By contracting with 
the stress-energy tensor we can construct the energy and angular momentum 
currents: ${\cal E}^{\alpha}=T^{\alpha}_{\beta}\xi^{\beta}=T^{\alpha}_{t}$, 
and ${\cal L}^{\alpha}=T^{\alpha}_{\beta}\chi^{\beta}=T^{\alpha}_{\phi}$, which 
are conserved, viz.,
\begin{equation}
{\cal L}^{\alpha}_{;\alpha}={\cal E}^{\alpha}_{;\alpha}=0.
\label{currents}
\end{equation}

Equations (\ref{continuity}), (\ref{ideal}), and (\ref{currents}) together with the 
homogeneous Maxwell equations admit 4 quantities that 
are conserved along magnetic flux surfaces,
the shape of which is given by a stream function $\Psi(r,\theta)$: 
The particle flux per unit magnetic flux,
\begin{equation}
\eta(\Psi)=\frac{\sqrt{-g}nu^r}{F_{\theta\phi}}=\frac{\sqrt{-g}nu^{\theta}}{F_{\phi r}}.
\label{eta}
\end{equation}
The angular velocity of magnetic field lines,
\begin{equation}
\Omega_F(\Psi)=F_{tr}/F_{r\phi}=F_{t\theta}/F_{\theta\phi}.
\label{Omega}
\end{equation}
The total energy and angular momentum per particle carried by the MHD flow,
\begin{eqnarray}
\label{E}
E(\Psi)=h u_t -\frac{\sqrt{-g}}{4\pi\eta}\Omega_F F^{r\theta},\\
\label{L}
L(\Psi)=-h u_{\phi}-\frac{\sqrt{-g}}{4\pi\eta} F^{r\theta}.
\end{eqnarray}
The above equations also yield the relation 
\begin{equation}
\frac{F_{r\theta}}{F_{\phi\theta}}=\frac{\Omega_F-v^{\phi}}{v^r},
\label{eq:Bphi/Br}
\end{equation}
where $v^{\phi}=u^{\phi}/u^{t}$, and $v^r=u^r/u^t$ are the corresponding components 
of the 3-velocity.  Equations (\ref{E}) and (\ref{L}) can be used to express 
$u^t$, $v^{\phi}$, and $F^{r\theta}$ in terms of $E$, $L$, and the Alfv\'en Mach number,
defined by $M^2=4\pi h\eta^2/n$.  One finds
\begin{eqnarray}
F^{r}_{\theta}= \frac{4\pi\eta}{\sin\theta}\frac{(g_{tt}+g_{t\phi}\Omega_F)L
+(g_{t\phi}+g_{\phi\phi}\Omega_F)E}{k_0 + M^2},\\
\label{u^t}
hu^t = h(-u_t + \beta u_{\phi})/\alpha^2 =(E-\Omega_F L)- \frac{M^2(E+\beta  L)}
{\alpha^2(k_0+M^2)},\\
\label{v^phi}
v^{\phi}=\frac{-\alpha^2 g_{\phi\phi}\Omega_F(E-\Omega_F L)
+M^2(g_{t\phi}E+g_{tt}L)}  {-\alpha^2 g_{\phi\phi}(E-\Omega_F L)
-M^2(g_{\phi\phi}E+g_{t\phi}L)},
\end{eqnarray}
where $k_0= g_{tt}+2g_{t\phi}\Omega_F + g_{\phi\phi}\Omega_F^2$.
Finally, using the normalization condition, $u^{\mu}u_{\mu}=-1$, we obtain for the 
poloidal velocity, defined as $u_p^2=u_ru^r+u_{\theta}u^{\theta}$, the relation
\begin{equation}
u_p^2+1=(\alpha u^t)^2-(u_{\phi}/\tilde{\omega})^2.
\label{u_p}
\end{equation}
Note that $u_p$ is the poloidal 4-velocity as measured by a ZAMO, and 
$\alpha u^t$ is the corresponding $t$ component of the 4-velocity. As seen,
the Lorentz factor of the inflow in the ZAMO frame approaches $\Gamma=\alpha u^t
\sim \alpha^{-1}$ on the event horizon.

Consider first the behavior of the solution near the horizon.  There $\alpha\rightarrow 0$
and so
\begin{equation}
v^{\phi}\rightarrow -\frac{(g_{t\phi}E+g_{tt}L)}{(g_{\phi\phi}E+g_{t\phi}L)}= \Omega_H.
\label{v_hi}
\end{equation}
The poloidal velocity is radial on the horizon, implying $u_p^2 \rightarrow  g_{rr}u^ru^r$. 
Using eq. (\ref{u_p}) we find,
\begin{equation}
v^r =u^r/u^t \rightarrow \alpha/\sqrt{g_{rr}} = -\Delta/(r^2+a^2).
\end{equation}
Substituting the above results into eq. (\ref{eq:Bphi/Br}), we finally obtain,
\begin{equation}
\frac{F_{r\theta}}{F_{\phi\theta}}= - \frac{r^2+a^2}{\Delta}(\Omega+\beta). 
\label{ratio1}
\end{equation}
Eq. (\ref{ratio1}) gives the boundary condition on the horizon.

The Boyer-Lindquist electric charge density in the flow is 
determined from Maxwell's equation: 
$F^{t\mu}_{\ \ ;\mu}=4\pi j^{t}$.  It can be readily shown that if the poloidal component 
of the magnetic field is radial on the event horizon (which is the case in the force-free
limit of an uncharged blck hole), then near the horizon the later equation reduces to,   

\begin{equation}
\frac{\partial}{\partial\theta}\left[\frac{\sin\theta}{\alpha^2}
(\Omega+\beta)F_{\phi \theta}\right]=
4\pi\sqrt{-g}j^{t}.\label{Max t2}
\label{Max-t}
\end{equation}

The poloidal current follows from the equation,
\begin{equation}
\frac{\partial}{\partial\theta}\left(\frac{\sin\theta}{\rho^2}
F_{r\theta}\right)=4\pi\sqrt{-g}j^r.
\label{Max-r}
\end{equation}
Using eqs. (\ref{eq:Bphi/Br}), (\ref{v_hi}), (\ref{Max-t}), and (\ref{Max-r}),
we finally recovers the result \citep{PC90})
\begin{equation}
j^r=\frac{\rho_e}{v_p},
\end{equation}
where $\rho_e=\alpha^2 j^t$, and $v_p=u_p/u_{\hat{t}}$ being the three 
velocity as measured by a ZAMO.
\end{document}